\documentclass[preprint, preprintnumbers, amsmath, amssymb, aps, prd, showkeys, nofootinbib, superscriptaddress]{revtex4-2}

\usepackage{amsmath, amssymb, parskip, xcolor, physics, natbib, graphicx, enumitem, lipsum, comment, amsfonts, setspace}

\usepackage[linkcolor=blue, colorlinks=true, citecolor=blue, urlcolor=blue]{hyperref}

\usepackage[mathscr]{eucal}

\usepackage{dcolumn}

\usepackage{setspace}

\def\equationautorefname~#1\null{(#1)\null}

\begin{document}


\title{Fractonic coset construction for spontaneously broken translations}

\author{Ameya Chavda}
\email{ameya.chavda@columbia.edu}
\affiliation{Center for Theoretical Physics, Department of Physics, Columbia University,
538 West 120th Street, New York, NY 10027, USA}

\author{Daniel Naegels}
\email{dn441@cam.ac.uk}
\affiliation{Center for Theoretical Physics, Department of Physics, Columbia University,
538 West 120th Street, New York, NY 10027, USA}
\affiliation{Department of Applied Mathematics and Theoretical Physics, University of Cambridge,
Wilberforce Road, Cambridge, CB3 0WA, United Kingdom}

\author{John Staunton}
\email{j.staunton@columbia.edu}
\affiliation{Center for Theoretical Physics, Department of Physics, Columbia University,
538 West 120th Street, New York, NY 10027, USA}

\date{\today}

\begin{abstract}
We study the homogeneous breaking of spatial translation symmetry concomitantly with the spontaneous breaking of other internal and spacetime symmetries, including dilations. We use the symmetry-breaking pattern as the only input to derive, via the coset construction, general effective field theories for the symmetry-originated modes associated with Goldstone's theorem, namely the Nambu-Goldstone candidates. Through explicit computations, we show that integrating out the explicit massive Nambu-Goldstone candidates or imposing symmetric constraints, namely the inverse Higgs constraints, to express massive modes in terms of the massless ones leads to physically distinct effective field theories. This sensitivity to the chosen method can be traced back to the homogeneous breaking of translations, the homogeneous aspect of the breaking induces a mixing between internal and spacetime symmetries at the level of the Lie algebra. This, in turn, leads to subtle discussions about the inverse Higgs constraints, in particular that they lead to a loss of generality in our specific examples. The derived general effective field theories also give rise to a broad class of theories exhibiting emergent enhanced shift symmetries, which constrain the mobility of the modes. The latter are referred to as fractonic modes.
\end{abstract}

\keywords{Goldstone physics, effective field theory, coset construction, inverse Higgs constraints, fractonic physics}

\maketitle

\begin{spacing}{1}
    \tableofcontents
\end{spacing}

\section{\label{sec:Introduction}Introduction}
One of the most powerful tools we have for understanding low-energy and nonrelativistic physics is Goldstone's theorem, which asserts that any theory with a continuous global symmetry group $G$ that is spontaneously broken to a subgroup $H$, where $H \subset G$, will necessarily include at least one gapless mode in its spectrum \cite{Goldstone:1962es}. Goldstone's theorem has a wide range of applicability for three main reasons. First, symmetries are broken everywhere in physics, including sitting in a room since this gives a preferred frame thereby breaking boosts. Second, one, in principle, only needs the symmetry-breaking pattern to derive the shape of an effective field theory. Third, the masslessness of the modes is a nonperturbative result. Examples of its applicability include magnons in ferromagnetic materials, phonons in superfluids,\footnote{For a pedagogical introduction to the relativistic superfluid, see \cite{Landry2019}.}  and kaon condensation in quantum chromodynamics. For a more comprehensive review, see \cite{Watanabe:2019xul, Brauner:2024juy}.

Recent progress in studying Nambu-Goldstone (NG) physics is in the spontaneous breaking of spacetime symmetries. This has, for example, yielded a deeper understanding of condensed matter systems as arising from the spontaneous breaking of Poincaré symmetry (e.g. \cite{Nicolis:2015sra}). Of the modes that arise from symmetry-breaking, not all of them will be massless. In this paper, we will use the term NG candidate to describe a fluctuation of the stable ground state in the direction of the NG candidates' associated broken-symmetry generator. If these fluctuations are independent and are massless as dictated by the symmetries of the theory, then they are termed NG modes.

One direction of research in NG physics has aimed to establish a counting rule for NG modes and develop methods for writing generic effective field theories (EFTs) while remaining as agnostic as possible about the high-energy theory for NG candidates and the way they could couple to matter fields. In general, the number of NG modes is less than the number of NG candidates, the latter of which we always take to be equal to the number of generators broken in the given symmetry-breaking pattern. As an example of when the number of NG candidates can be reduced, let us consider the following Lagrangian written in terms of NG candidates $\pi^1$ and $\pi^2$:
\begin{equation}
\label{eq:ConjugationReduction}
\mathcal{L} = \frac{1}{2}\left(\partial_t \pi^1\right)^2 + \frac{1}{2}\left(\partial_t \pi^2\right)^2 + \frac{1}{2}\lambda \pi^1 \partial_t \pi^2 - \frac{1}{2}\lambda \pi^2 \partial_t \pi^1 - ...,
\end{equation}
where the ellipsis indicates additional terms involving spatial derivatives of $\pi^1$ and $\pi^2$. Here, $\pi^1$ and $\pi^2$ are canonically conjugate, so at low energies below any mass scale, they only form 1 independent degree of freedom. This constitutes a \textit{dynamical reduction} of the number of NG modes compared to the number of possible NG candidates \cite{Watanabe:2019xul}. A counting rule for the breaking of internal symmetries, i.e. symmetries that commute with the Poincar\'{e} group, where the reduction of NG modes is solely dynamical, has been established in \cite{Watanabe:2011ec, Watanabe:2012hr} by writing the most general effective field theory.\footnote{For example, when there is unbroken Poincar\'{e} invariance and only internal symmetries are broken, the number of NG modes is equal to the number of broken internal symmetry generators.}

\begin{figure}
\label{fig:String}
\includegraphics[scale=0.5]{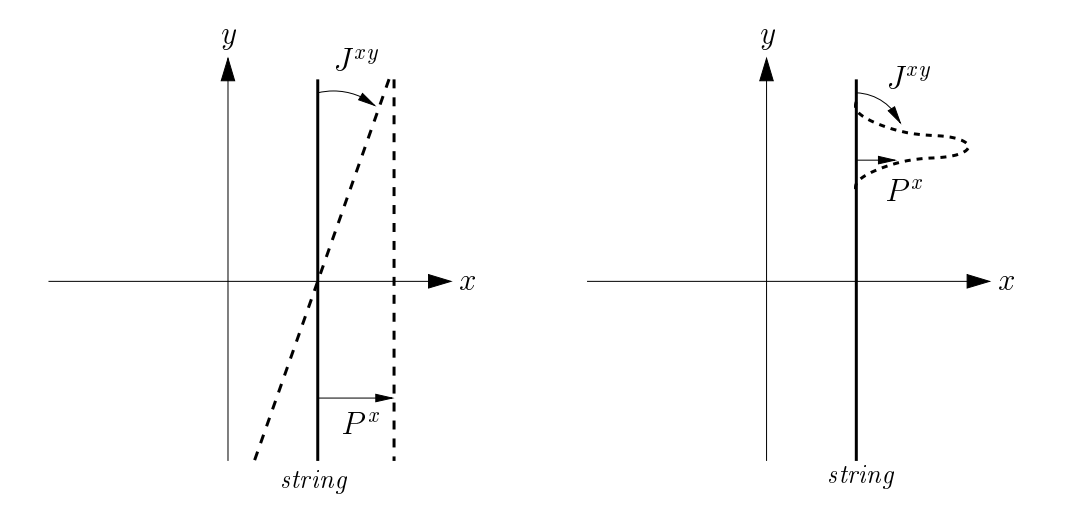}
\caption{Reproduced from \cite[Fig. 1]{Low:2001bw}. The solid line is a string that breaks (2+1)-dimensional Poincar\'{e} invariance to a two-dimensional subspace. Globally, as shown on the left, $x$-translations, generated by $P^x$, and rotations, generated by $J^{xy}$, are not equivalent. Locally, however, translations and rotations can be made to be the same, as shown on the right.}
\end{figure}

There are also nondynamical mechanisms for reducing the number of NG modes relative to the number of NG candidates when considering the breaking of spacetime symmetries, i.e. symmetries that act on spacetime coordinates. For instance, let us consider the scenario depicted in \cite[Fig. 1]{Low:2001bw}, reproduced in \hyperref[fig:String]{Figure 1}, where a string extends along the $y$-axis. Globally, a rotation and a translation on the string is clearly different; but, at a specific point on the string, a local translation can be equivalent to a local rotation. This leads to an additional reduction in the number of NG candidates since the fluctuations along those two symmetries are redundant meaning their NG candidates are also the same. To this point, there is no counting rule for the case of spontaneously broken spacetime symmetries.

That being said, in the context of effective field theories, there have been a number of methods developed to identify the appropriate NG modes from the NG candidates. The most well known is the inverse Higgs constraint, which allows one to relate one field to the derivative of another field. The current paradigm is that the procedure of imposing inverse Higgs constraints is equivalent to integrating out the massive modes or eliminating redundant NG candidates when spacetime translations are unbroken \cite{Brauner:2014aha, Klein:2017npd, Finelli:2019keo}. One of the goals of this paper is to further explore the consequences of inverse Higgs constraints when spacetime translations \textit{are} broken through the use of the coset construction. A review of the coset construction for broken spacetime symmetries is in \hyperref[app:CosetConstruction]{Appendix A}. In \hyperref[app:IHCReview]{Appendix B}, we provide an overview of the current state of the art for inverse Higgs constraints. In order to study this question, we would like our symmetry-breaking pattern to include translations, rotations, and dilations. Breaking translations puts us in a situation where we can study subtleties involved with inverse Higgs constraints. Breaking rotations not only gives us a possible redundancy with translations, but it is also common in physics, such as in crystals. Lastly, dilation symmetry often appears near critical points in statistical physics, so we are motivated to at least include it in our chosen symmetry-breaking patterns.

To be more specific, we provide two interesting cases studies, inspired by the toy model in \cite{Argurio:2021opr}, that spontaneously break translations, dilations, and rotations: the metafluid and the helical superfluid. The exact symmetry-breaking patterns will be described in more depth in \autoref{s:MF} and \autoref{s:HSF} respectively. In both cases, we break spatial translations homogeneously, meaning that there will be an unbroken translation generator after symmetry-breaking that is a combination of broken translations and some other broken internal generator. This is more tractable compared to pure translation breaking since, in the latter case, there is difficulty in parametrizing generic perturbations in the broken directions \cite{Brauner:2024juy}. The breaking of spacetime symmetries is not only interesting for the further development of NG physics, but also widely applicable since any physical state of matters breaks some spacetime symmetries. For example, a point particle spontaneously breaks translations and boosts \cite{Delacretaz:2014oxa}. The consideration of homogeneous breaking is particularly relevant in condensed matter theory, as the crystal structure can be approximated as smooth and homogeneous at low energies, such as in solids studied in \cite{Nicolis:2013lma} and \cite{Nicolis:2015sra}. The first goal of this paper will be to utilize these case studies to elucidate many subtleties having to do with inverse Higgs constraints.

In the process of studying the metafluid and helical superfluid, we will encounter a large class of theories that have \textit{fractonic modes} in their spectrum. Fractonic modes were first encountered in lattice models \cite{Vijay:2015mka,Vijay:2016phm} and are characterized by restricted mobility. By restricted mobility, we mean that these modes cannot move in certain directions on their own and, instead, need interactions. Fractonic modes were then studied in continuous field theories, where a signature for the appearance of fractonic modes is the conservation of multipole moments associated to polynomial shift symmetries \cite{Pretko:2016kxt, Gromov:2018nbv}. In this paper, we define a fractonic mode to be a particle whose dispersion relation gives $0$ frequency in some or all directions. Three examples of fractonic dispersion relations are
\begin{equation}
\omega^2 = 0, \qquad \omega^2 = p_x p_y, \qquad \omega^2 = p_x^2.
\end{equation}
In the first case, we have pure fractonic behavior. In the second case, in the $x$ or $y$ direction, where $p_y = 0$ or $p_x = 0$ respectively, $\omega$ vanishes. In the last example, $\omega$ vanishes in the $y$ direction, where $p_x = 0$. This is a commonly used definition of fractonic behavior (e.g. \cite{Pretko:2018jbi, Seiberg:2019vrp,Seiberg:2020bhn}). As a final note, if the fractonic mode is immobile, it is a fracton, if it is able to move in one spatial direction, it is called a lineon, and if it is able to move in two spatial directions, it is called a planeon. A useful review of fractonic physics is in \cite{Grosvenor:2021hkn}.

Fractonic physics is itself an interesting pursuit both practically and conceptually. Practically, they can be found in crystal defects captured by elasticity theory \cite{Pretko_2018,PhysRevB.100.134113}. It has also been applied in the context of hydrodynamics, where polynomial symmetry is taken as an assumption in the studied models (e.g. \cite{Gromov:2020yoc,Grosvenor:2021rrt,Glodkowski:2022xje}) and is sometimes broken leading to the fractonic superfluid (e.g. \cite{Yuan:2019geh,Chen:2020jew,Stahl:2021sgi,Stahl:2023prt,Afxonidis:2023tup,Glodkowski:2024ova}). Conceptually, the dispersion relations demonstrate a mixing of UV and IR physics since one could go to arbitrarily high momenta, $p \rightarrow \infty$, in some directions, but still have zero energy, $\omega = 0$. Second, it is also common in fractonic physics to end up with discontinuous fields or fields without properly defined Fourier transforms. This can intuitively be seen from the dispersion relations since the behavior at low-energy can be affected by fields at infinitely high momenta \cite{Seiberg:2020bhn}.

The second goal of this paper will be to study the emergence of enhanced shift symmetry from a symmetry-breaking pattern that does not include any enhanced shift symmetry. By enhanced shift symmetry, we mean that the theory is invariant under transformations of the field by a shift of a function of arbitrary spatial dependence, a generalization of polynomial shifts. Note that this is different than the hydrodynamics situation which explicitly incorporates polynomial shifts into the symmetry-breaking pattern. We will do this using the metafluid and helical superfluid, whose spectra are already known to have fractonic modes in some cases \cite{Argurio:2021opr}. Since fractonic behavior only appears in the absence of interactions, we will only study these theories up to quadratic order in the fields. We are not the first ones to study fractonic physics with the coset construction (see for example \cite{Pena-Benitez:2021ipo, Hirono:2021lmd}). In particular, \cite{Hirono:2021lmd} spontaneously break the Galilean group and an additional U(1) for the supersolid. By introducing topological defects into the obtained EFT, they can describe dislocations and disclinations, which shows that their fractonic behavior can be traced back to the semidirect product structure of the spatial translations and spatial rotation groups. In the present paper, we instead look for fractonic modes characterized by their dispersion relations as opposed to their topological nature. In addition, while the symmetry-breaking pattern for the metafluid is similar to the one studied in \cite{Hirono:2021lmd}, the helical superfluid symmetry-breaking pattern is different. Lastly, while \cite{Hirono:2021lmd} has emergent charge and dipole conservation, we explore the possibility of general enhanced shift symmetries that are not necessarily polynomial.

As a road map for the paper, in \autoref{sec:ToyModel}, we review the 
toy model in \cite{Argurio:2021opr} and then extract the symmetry-breaking pattern that we later use to derive wider classes of effective field theories. These classes are based on two different symmetry-breaking patterns, which we call the metafluid and the helical superfluid. Then, in \autoref{s:MF}, we conduct a detailed coset construction for the metafluid, showing that imposing inverse Higgs constraints results in conditions that cannot be satisfied when fields are expressed with well-defined Fourier modes. We further show that, on the other hand, there is a well-defined theory upon integrating out the massive modes. We demonstrate that a large subclass of those effective field theories exhibit enhanced shift symmetry and result in fractonic dispersion relations. In \autoref{s:HSF}, we extend our analysis to the helical superfluid, deriving two distinct effective field theories. One arises from the imposition of inverse Higgs constraints and the other from integrating out the massive modes without imposing inverse Higgs constraints. Once again, we explore the conditions under which the resulting effective field theories exhibit emergent enhanced shift symmetries and fractonic dispersion relations. We discuss our results further in \autoref{s:Discussion}. In \autoref{s:MF} and \autoref{s:HSF}, we do not claim to derive the most general possible effective field theory. Indeed, when spacetime symmetries are broken, the choice of parametrization might lead to a loss of generality. Moreover, we choose a specific representation of the symmetry group $G$, but it is not clear that our chosen representation is the most general representation of the group that exists. Lastly, we use the Maurer-Cartan form to yield building blocks for the effective field theory, but that does not mean the most general effective field theory is the one built out of the Maurer-Cartan coefficients. 

\textit{Summary of results:} Using two different symmetry-breaking patterns, we do the following:
\begin{itemize}
\item Demonstrate that the use of inverse Higgs constraints when translations are homogeneously broken can lead to a loss of generality in the resulting low-energy effective field theories.
\item Show that a large subset of these effective field theories have emergent enhanced shift symmetries and associated fractonic dispersion relations.
\item Provide a code that can perform the coset construction for generically complicated symmetry-breaking patterns \cite{SuppMat}.
\end{itemize}

\textit{Notation and conventions:} Throughout the paper, we will work in $2 + 1$ dimensional Minkowski spacetime and use the mostly plus signature $\left(-, +, +\right)$. The Fourier transform conventions are as follows:
\begin{equation}
\tilde{f}\left(k\right) = \int f\left(x\right) e^{-ikx} \dd[n]{x} \qquad \text{and} \qquad f\left(x\right) = \int \tilde{f}\left(k\right) e^{ikx} \frac{\dd[n]{k}}{\left(2\pi\right)^n}.
\end{equation}
The rotation matrix is defined as:
\begin{equation}
    R^i_{\hphantom{i}j} \left(\theta\right) = \begin{pmatrix}
    \cos\theta & -\sin\theta \\ \sin\theta & \cos\theta
\end{pmatrix}.
\end{equation}
In this paper, we never consider the explicit breaking of symmetries and so we use broken symmetries and spontaneously broken symmetries interchangeably. 

\section{\label{sec:ToyModel} A Toy Model}
The symmetry-breaking patterns are based on a nonrelativistic toy Lagrangian that, after spontaneous symmetry-breaking, successfully predicts low-energy theories that exhibit emergent enhanced shift symmetry, which in turn leads to fractonic behavior \cite{Argurio:2021opr}. In this section, we briefly review the low-energy theories whose theory prior to symmetry-breaking is given by
\begin{equation}
\label{eq:ToyModelL}
    \mathcal{L} = \partial_t \Phi^* \partial_t \Phi + A \partial_i \Phi^* \partial_i \Phi + \frac{1}{2} \partial_t \Xi \partial_t \Xi - \frac{1}{2} \partial_i \Xi \partial_i \Xi - B \frac{(\partial_i \Phi^* \partial_i \Phi)^2}{\Xi^6} - \frac{A^2}{4B} \Xi^6,
\end{equation}
where $\Phi$ is a complex scalar field, $\Xi$ is a real compensator field used to enforce scale invariance, and $A$ and $B$ are positive constants. Intuitively, the fact that $A$ is positive means the gradient terms are of the wrong sign for typical Lorentz invariant theories, akin to theories with the wrong sign for mass terms in theories with the standard Mexican hat potential. The quartic gradient plays the role of the quartic term in the usual Mexican hat potential, but for the gradient instead allowing for the breaking of translation symmetry. Finally, the compensator field, $\Xi$, both enforces scale invariance as well as helps ensure there are nontrivial ground states that minimize the energy and break a number of spacetime symmetries.

The details of the computation are further discussed in \cite{Argurio:2021opr}. Here, we review the symmetries of the Lagrangian and describe the two ground states that break some of those symmetries. Specifically, the symmetries of the Lagrangian are:
\begin{enumerate}[label=\roman*.]
    \item U(1), with generator $Q$ and parameter $\alpha$:
        \begin{equation}
        \label{eq:FirstSym}
            \Phi \rightarrow e^{i \alpha} \Phi \qquad \text{and} \qquad \Xi \rightarrow \Xi.
        \end{equation} 
    \item Real and Imaginary Shifts, with generators $S_i = \left(S_1, S_2\right)$ and parameters $s^i$:
        \begin{equation}
            \Phi \rightarrow \Phi + s^1 + i s^2 \qquad \text{and} \qquad \Xi \rightarrow \Xi.
        \end{equation}
    \item Dilations, with generator $D$ and parameter $\lambda$:
        \begin{equation}
            x^\mu \rightarrow e^{-\lambda} x^\mu, \qquad \Phi \rightarrow e^{\Delta \lambda} \Phi, \qquad \text{and} \qquad \Xi \rightarrow e^{\Delta \lambda} \Xi,
        \end{equation}
        where $\Delta = 1/2$ for the given toy model.
    \item Rotations, with generator $J$ and parameter $\theta$:
        \begin{equation}
            x'^i \rightarrow R^i_{\hphantom{i}j}\left(\theta\right) x^j \qquad \text{and} \qquad \Phi'\left(x'\right) \rightarrow \Phi\left(x\right).
        \end{equation}
    \item Spacetime translations, with generator $P_\mu$ and parameter $a^\mu$:
        \begin{equation}
        \label{eq:LastSym}
            x'^\mu \rightarrow x^\mu + a^\mu \qquad \text{and} \qquad \Phi'\left(x'\right) \rightarrow \Phi\left(x\right).
        \end{equation}
\end{enumerate}
We note that while the theory does contain a complex shift symmetry, it does not contain enhanced shift symmetry as described by \cite{Pretko:2018jbi}. In other words, the shifts are by constant values, rather than arbitrary functions of position. This distinction will become important once we describe the low-energy theories for the NG modes.

The authors of \cite{Argurio:2021opr} seek to identify ground states that break translations homogeneously. By homogeneous translation breaking, we mean that the perturbative theory after spontaneous symmetry-breaking does not explicitly depend on spacetime coordinates. At the level of the algebra, this means that the breaking of translations is compensated by the breaking of an additional internal symmetry to ensure that a diagonal combination of the two broken generators remains unbroken. We will denote unbroken combinations of generators with a bar on top, such as $\bar{P}$. In this case, there are two such ground states:
\begin{enumerate}[label=\roman*.]
    \item The metafluid:
        \begin{equation}
            \Phi(t,x,y) = b(x+iy) \qquad \text{and} \qquad \Xi(t,x,y) = v,
        \end{equation}
        where $v$ and $b$ are real and complex constants, respectively.\footnote{In general, the real and imaginary components of $\Phi$ could have different coefficients. To keep rotational symmetry, we choose the ground state that has the same factor $b$.} The translation and rotation generators are preserved diagonally in the following sense: 
        \begin{equation}
            \bar{P}_i = P_i + S_i \qquad \text{and} \qquad \bar{J} = J + Q.
        \end{equation}
        So that the broken ($X_a$) and unbroken ($T_A$) generators excluding the unbroken translations are, respectively: 
        \begin{equation}
        \label{eq:MFdecomp}
            X_a = \left\{Q, S_i, D\right\} \qquad \text{and} \qquad T_{A} = \left\{\bar{J}\right\}.
        \end{equation}

    \item The helical superfluid:
        \begin{equation}
        \label{eq:HSFGroundState}
            \Phi(t,x,y) = \rho e^{ikx} \qquad \text{and} \qquad \Xi(t,x,y) = v,
        \end{equation}
        where $\rho$, $k$, and $v$ are real constants. For this ground state, the unbroken translation generator is: 
        \begin{equation}
            \bar{P}_\mu = P_\mu + k\delta_\mu^{\hphantom{\mu}1}Q.
        \end{equation}
        As such, the broken and unbroken generators excluding the unbroken translations are:
        \begin{equation} 
        \label{HSFgen}
            X_a = \left\{Q, S_i, D, J\right\} \qquad \text{and} \qquad T_{A} = \varnothing.
        \end{equation}
\end{enumerate}
In both cases, the authors of \cite{Argurio:2021opr} expand the fields perturbatively around the ground state to quadratic order in the fields and obtained two massless modes and one massive mode. Then, they integrate out the massive mode to obtain the respective effective field theories for the remaining massless NG modes. Explicitly, for the metafluid, this means expanding the fields in the following way:
\begin{equation}
    \Phi = b (x + i y) + b \left[u^1(t, x, y) + i u^2(t, x, y) \right]
\end{equation}
and 
\begin{equation}
    \Xi = \nu + \tau(x, y, z).
\end{equation} 
In this case, the $\tau$ mode is massive and needed to be integrated out. The resulting EFT for $u^i$ up to fourth order in derivatives is
\begin{equation}
\label{eq:ArgurioMF}
\mathcal{L}\left(u^i\right) = \abs{b}^2 \partial_t u_i \partial_t u^i - \frac{v^2}{72}\partial_\mu \partial_i u_j \partial^\mu \partial^j u^i.
\end{equation}
Note that in \cite{Argurio:2021opr}, the authors are able to write down a Lagrangian up to sixth order in derivatives, but we will not go that far in the derivative expansion in this paper. The resulting (massless) dispersion relations at low momentum are then found to be, with $\vb{p}$ being the spatial momentum,
\begin{align}
\omega_1^2 &= 0, \\
\omega_2^2 &= \frac{v^2}{72\abs{b}^2} \vb{p}^4 + \mathcal{O}\left(p^6\right).
\end{align}

Notice that there is one trivial mode, which is identified as a pure fractonic mode.\footnote{The trivial mode appears without any corrections of the form $\mathcal{O}(p^n)$, meaning $\omega = 0$ in the free theory is valid at any momentum.} As noted in \cite{Argurio:2021opr}, the fact that this fractonic mode appears is coupled with the emergence of an enhanced shift symmetry in \autoref{eq:ArgurioMF}:
\begin{equation}
u^i \rightarrow u^i + a^i + b^i_{\hphantom{i}j} x^j + c^i_{\hphantom{i}jk} x^j x^k,
\end{equation}
where $a^i$, $b^i_{\hphantom{i}j}$, and $c^i_{\hphantom{i}jk}$ are constant tensors. The invariance under transformations up to quadratic order implies the conservation of the charge associated with the usual constant shift symmetry and also higher-order multipole moments, in this case dipole and quadrupole moments. It has been shown that conservation of multipole moments and enhanced shift symmetries are a possible signature of fractonic behavior at the level of the dispersion relations \cite{Pretko:2018jbi, Seiberg:2019vrp}. Although we present the truncated Lagrangian up to the fourth order in derivatives, it should be noted that in the original work, the authors demonstrate that this symmetry holds even up to the sixth order in derivatives.

In \cite{Argurio:2021opr}, the authors perform an analogous procedure for the helical superfluid by first expanding around the ground state in \autoref{eq:HSFGroundState} as follows:
\begin{equation}
    \Phi(t,x,y) = \rho e^{ikx} \left[1 + \phi(t,x,y) \right] = \rho e^{ikx} \left[1 + \phi_R\left(t, x, y\right) + i \phi_I\left(t, x, y\right)\right]
\end{equation}
and 
\begin{equation}
    \Xi(t,x,y) = \nu \left[ 1 + \tau(t,x,y) \right].
\end{equation}
Once again, the resulting theory contains a massive mode that is a combination of $\tau$ and $\phi_R$. By redefining these two fields in terms of one massive field $\eta$ and one massless field $\varphi$ and integrating out the massive field $\eta$, they find the following theory, which we quote up to second-order in fields and derivatives:
\begin{equation}
\mathcal{L}\left(\varphi, \phi_I\right) = \rho^2\left(\partial_t \phi_I\right)^2 + \frac{1}{2}\left(\partial_t \varphi\right)^2 - \frac{v^2}{2\left(v^2 + 18\rho^2\right)}\partial_i\varphi \partial^i \varphi.
\end{equation}
In the original calculation, \cite{Argurio:2021opr} writes down this Lagrangian up to fourth order in derivatives. The resulting dispersion relations, up to only quadratic order, are
\begin{align}
\omega_1^2 &= 0, \\
\omega_2^2 &= \frac{v^2}{18\rho^2 + v^2}\vb{p}^2 + \mathcal{O}(p^4).
\end{align}
As before, the presence of fractonic behavior in the dispersion relations can be traced back to the fact that the $\mathcal{L}$ is symmetric under the following enhanced shift symmetry: 
\begin{equation}
\phi_I \rightarrow \phi_I + f(x, y).
\end{equation}
In the theory expanded to fourth order in derivatives, this symmetry is slightly modified to be a symmetry up to quadratic order in $x$, but still retains full symmetry under shifts in any function of $y$. See \cite[eq. 2.60]{Argurio:2021opr} for full details. 

As discussed in \hyperref[app:CosetConstruction]{Appendix A}, NG modes transform nonlinearly and often have a shift symmetry. Here, we would like to use the generality of the coset construction to see if that shift symmetry can become an enhanced shift symmetry, which is often associated to fractonic behavior in the dispersion relations, in models that break spacetime symmetries. To start, let us establish the algebra associated to the individual symmetry transformations in \autoref{eq:FirstSym} -- \autoref{eq:LastSym}, under which a generic field $\phi\left(x\right)$ transforms as:
\begin{equation}
\phi'\left(x'\right) = e^{i\alpha}\left(e^{\Delta \chi}\phi\left(x\right) + \sigma_R + i\sigma_I\right) \quad \text{and} \quad x'^\mu = e^{-\chi}R^\mu_{\hphantom{\mu}\nu}\left(x^\nu + a^\nu\right).
\end{equation}
Once we know how the field and coordinates transform, we can systematically derive the Lie algebra for the symmetry group of the theory. This procedure closely follows the computation of the Poincar\'{e} Lie algebra presented in \cite{Weinberg1}. First, let us rewrite the above transformation as the action of a representation, $D$, of a symmetry group element $g$:
\begin{equation}
D\left[g\left(\alpha, \sigma^1, \sigma^2, \chi, \theta, a\right)\right]\phi\left(x\right) = e^{i\alpha}\left(e^{\Delta \chi}\phi\left(e^{\chi} R^{-1}\left(\theta\right) x - a\right) + \sigma^1 + i\sigma^2\right).
\end{equation}
The parameters of the group element were defined in \autoref{eq:FirstSym} -- \autoref{eq:LastSym}. The most subtle commutator to derive is the one between shifts and dilations, since it involves a mixing of field transformations and spacetime transformations. We compute the following group transformation rule by studying how the equivalent representation acts on the field $\phi$:
\begin{equation}
g\left(0, \sigma^1, 0, 0, 0, 0\right)g\left(0, 0, 0, \chi, 0, 0\right)g^{-1}\left(0, \sigma^1, 0, 0, 0, 0\right) = g\left(0, \left(1 - e^{\Delta \chi}\right)\sigma^1, 0, \chi, 0, 0\right).
\end{equation}
Taking $g\left(t\right) = e^{itT}$, where $T$ is the generator and $t$ is the parameter, the left-hand side expands to $1 + i\chi D - \sigma^1 \chi \comm{S_1}{D}$, while the right-hand side expands to
\begin{equation}
g\left(0, \left(1 - e^{\Delta \chi}\right)\sigma^1, 0, \chi, 0, 0\right) \approx 1 + i\chi D - \Delta\chi\sigma^1 S_1.
\end{equation}
Matching both sides, we find that the commutator between $S_1$ and $D$ is
\begin{equation}
\comm{S_1}{D} = i\Delta S_1.
\end{equation}
\vspace{-1.25cm}

Following similar steps for the rest of the generators, we find that the nontrivial commutators of the symmetry Lie Algebra are:
\begin{equation}
\label{eq:FullLieAlg}
\begin{aligned}
\comm{Q}{S_i} &= -i\epsilon_i^{\hphantom{i}j}S_j \qquad & \qquad \comm{S_i}{D} &= i \Delta S_i \\
\comm{D}{P_\mu} &= iP_\mu \qquad & \qquad \comm{J}{P_i} &= -i\epsilon_i^{\hphantom{i}j} P_j.
\end{aligned}
\end{equation}
While they can be derived rigorously, they can also be understood intuitively. The commutator of $Q$ and $S_i$ is analogous to the commutator of $J$ and $P_i$, except in the target space, i.e. the 2D space spanned by the real and complex parts of $\phi$. Similarly, the commutator of $S_i$ and $D$ is analogous to the commutator of $D$ and $P_\mu$. The algebra's parameter, $\Delta$, tells us that a consistent representation of this algebra will be such that the field transforming under shifts will have a conformal weight $\Delta$ in the same way the commutator of $D$ and $P_\mu$ tells us that the coordinates have a conformal weight of $1$. In the next two sections, we will utilize this algebra as our starting point for performing the coset construction.

\section{\label{s:MF} The Metafluid}
Recall that the metafluid is described by the NG modes that result from the symmetry-breaking pattern in \autoref{eq:MFdecomp}. The nonzero elements of the Lie algebra in terms of those generators are
\begin{equation}
\label{eq:MFLieAlgebra}
    \begin{aligned}
        \comm{Q}{S_i} &= -i\epsilon_i^{\hphantom{i}j} S_j, \quad & \quad \comm{S_i}{D} &= i\Delta S_i, \quad & \quad \comm{D}{\bar{P}_\mu} &= i\left(\bar{P}_\mu - \left(\Delta + 1\right)\delta_\mu^{\hphantom{\mu}i} S_i\right), \\
        \comm{Q}{\bar{P}_i} &= -i\epsilon_i^{\hphantom{i}j} S_j, \quad & \quad \comm{S_i}{\bar{J}\:} &= i\epsilon_i^{\hphantom{i}j} S_j, \quad & \quad \comm{\bar{J}}{\bar{P}_i} &= -i\epsilon_i^{\hphantom{i}j} \bar{P}_j.
    \end{aligned}
\end{equation}
Our goal in this section is to construct as general a Lagrangian as possible for massless NG candidates that arise from this symmetry-breaking pattern. Since each of the broken generators and unbroken translations are multiplets of the invariant subgroup $H_0 =$ SO(2), where $H_0$ is the subgroup $H$ without the unbroken translations, we can do this with the coset construction. As noted in the introduction, we will possibly lose some generality at two stages in the calculation. First, since we are breaking spacetime symmetries, it is not guaranteed that different coset parametrizations will yield the same theory \citep{McArthur:2010zm, Creminelli:2014zxa, Finelli:2019keo}. Second, we will make use of the Maurer-Cartan form which nicely gives us objects that are covariant under the full symmetry group. There is no guarantee that this will yield \textit{all} possible operators invariant under the transformations of the fields. For completeness, we derive all the transformation laws in \hyperref[app:TransformationLaws]{Appendix D}. With that said, we will still obtain a sufficiently general theory to talk about the emergence of enhanced shift symmetry and fractonic behavior as well as an interesting case study in the application of inverse Higgs constraints.

To set the stage, we will define our coset parametrization as
\begin{equation}
\label{eq:MFCosetparametrization}
U\left(x, \sigma^i, \pi, \chi\right) = e^{ix^\mu \bar{P}_\mu} e^{i\sigma^i S_i} e^{i\pi Q} e^{i\chi D}.
\end{equation}
As suggested by \cite{Klein:2017npd}, a choice of parametrization that includes the most possible exponential factors is the most suitable to impose inverse Higgs constraints. Since $\comm{S_i}{S_j} = 0$, it does not matter that we combine the components of $S_i$ into one exponential.

With a chosen coset parametrization, we will be able to write down the Maurer-Cartan form. To make the computations simpler, let us note that $\bar{P}_i = P_i + k S_i$. As a result, we can let $\psi^i \equiv \sigma^i + k x^i$ and write
\begin{equation}
\label{eq:EquivMFparametrization}
U\left(x, \psi^i, \pi, \chi\right) = e^{i x^\mu P_\mu} e^{i\psi^i S_i} e^{i\pi Q} e^{i\chi D}.
\end{equation}
To recover any results on the true NG candidates, $\sigma^i$, we simply need to undo this field redefinition. Since both expressions for $U$ are equal, there are no subtleties that can arise from different coset paramtereizations.\footnote{We also computed the transformation laws and Maurer-Cartan forms directly using $\sigma^i$ and $\bar{P}_\mu$ and found the same results. The method we present here is computationally simpler.} However, it is important to remember that $\psi^i$ is not a small field the same way $\sigma^i$ is. In particular, since we will eventually want to expand in derivatives, we must do it in terms of $\sigma^i$ since $\partial \psi^i$ will contain an additional constant that would mix up the order of derivatives in the expansion. Additionally, the coefficients of the Maurer-Cartan form are determined using $\bar{P}_\mu$ instead of $P_\mu$. With that said, after some immediate simplifications from commutators that are just $0$, the Maurer-Cartan form becomes
\begin{equation}
U^{-1} \partial_\mu U = i\left(e^{-i\chi D} P_\mu e^{i\chi D} + e^{-i\chi D} e^{-i\pi Q} \partial_\mu \psi^i S_i e^{i\pi Q} e^{i\chi D} + \partial_\mu \pi Q + \partial_\mu \chi D\right).
\end{equation}
Making use of the results in \autoref{eq:MChelp1} and \autoref{eq:MChelp2}, we can write
\begin{equation}
U^{-1} \partial_\mu U = i\left(e^{\chi} \bar{P}_\mu + \left(e^{-\Delta \chi} R^i_{\hphantom{i}j}\left(\pi\right)\partial_\mu \psi^j - e^{\chi} \delta_\mu^{\hphantom{\mu}i}\right)S_i + \partial_\mu \pi Q + \partial_\mu \chi D\right),
\end{equation}
where we used the fact that the transpose of a rotation is equal to its inverse and we rewrote the broken translation generator, $P_\mu$, as the unbroken translation generator, $\bar{P}_\mu$.

We are now in a position to write down the Maurer-Cartan coefficients, $\varpi_\mu^{\hphantom{\mu}a}$, and the coset tetrad, $e_\mu^{\hphantom{\mu}\nu}$, as defined in \autoref{eq:GenericMCForm}:
\begin{equation}
\label{eq:MFBuildingBlocks}
    \begin{aligned}
        e_\nu^{\hphantom{\nu}\mu} &= e^{\chi} \delta_\nu^{\hphantom{\nu}\mu}, \quad & \quad \left(\varpi^Q\right)_\mu &= e^{-\chi} \partial_\mu \pi, \\
        \left(\varpi^S\right)_\mu^{\hphantom{\mu}i} &= e^{-\left(\Delta + 1\right)\chi} R^i_{\hphantom{i}j}\left(\pi\right)\partial_\mu \psi^j - \delta_\mu^{\hphantom{\mu}i}, \quad & \quad \left(\varpi^D\right)_\mu &= e^{-\chi} \partial_\mu \chi.
    \end{aligned}
\end{equation}
Note that each of the Maurer-Cartan coefficients are already at first order in the fields, which can be checked by taking the small fields $\sigma^i$, $\pi$, and $\chi$ to $0$ and finding that the Maurer-Cartan form is also $0$. The coefficient of the unbroken translation generator provides a natural definition of an invariant measure,
\begin{equation}
    \det\left(e\right)\dd[3]{x} = e^{3\chi} \dd[3]{x}.
\end{equation}
Note that, since $\sigma^i$ transforms as a vector under rotations, as shown in \autoref{eq:MFRottrans}, the Maurer-Cartan coefficients can be classified by how they transform under unbroken rotations:
\begin{equation}
    \begin{aligned}
        \text{Scalars: } & \left(\varpi^Q\right)_0, \: \: \left(\varpi^D\right)_0,\\
        \text{Vectors: } & \left(\varpi^S\right)_0^{\hphantom{0}i}, \: \: \left(\varpi^Q\right)_i, \: \: \left(\varpi^D\right)_i, \\
        \text{Tensor: } & \left(\varpi^S\right)_i^{\hphantom{i}j}.
    \end{aligned}
\end{equation}
In order to be invariant under the full symmetry group, we must take scalar combinations of these Maurer-Cartan coefficients by appropriately contracting the spatial indices. The final building block one would need to construct an invariant action is the covariant derivative. By \autoref{eq:GeneralCovD}, this is
\begin{equation}
    \nabla_\mu = e^{-\chi} \partial_\mu. 
\end{equation}

The Maurer-Cartan coefficients along with the covariant derivative and the invariant integration measure constitute the building blocks for an invariant action. We would want an action with only \textit{a priori} massless NG candidates, i.e. NG modes. Following the logic in \autoref{app:IHCReview}, we can examine the algebra in \hyperref[eq:MFLieAlgebra]{Appendix B} and predict that there will be two mass terms in the Lagrangian, as there are two broken generators $X_a = \left\{Q, D\right\}$ for which there exists a commutator $\comm{X_a}{\bar{P}_\mu} \supset X_b$\footnote{Since there are four broken generators, counting the components of $S_i$ as distinct generators, there should be \textit{a priori} two massless modes remaining, $\sigma^i$.}. The standard way to do reduce this Lagrangian to one only involving massless modes is the inverse Higgs mechanism. However, as we will see in \autoref{s:MFIHC}, this will lead to inconsistencies and unphysical theories. Instead, we will work directly with these building blocks as they are and integrate out any massive modes we encounter at the end of the calculation. This will be covered in the second subsection, \autoref{s:MFNoIHC}.

\subsection{\label{s:MFIHC} Imposing the inverse Higgs constraints}
The standard way to eliminate massive NG candidates is to impose inverse Higgs constraints. In order to do this (see \hyperref[app:IHCReview]{Appendix B}), we first need to isolate the commutators involving unbroken translations and two broken generators. Those commutators, from \autoref{eq:MFLieAlgebra}, are
\begin{equation}
\comm{Q}{\bar{P}_i} = -i\epsilon_i^{\hphantom{i}j} S_j \qquad \text{and} \qquad \comm{D}{\bar{P}_i} = i\left(\bar{P}_i - \left(\Delta + 1\right)S_i\right).
\end{equation}
There are a total of four commutators here, since $\bar{P}_i$ has two components. As discussed in \hyperref[app:IHCReview]{Appendix B}, inverse Higgs constraints set particular Maurer-Cartan coefficients to $0$ based on these generators. In general, for two broken generators $X_a$ and $X_b$, if $\comm{X_a}{\bar{P}_i} \supset X_b$, then the NG candidate associated to $X_a$ can be expressed as a derivative of the NG candidate associated to $X_b$, so long as imposing this constraint does not violate any of the unbroken symmetries contained in $H_0$. This is accomplished by setting 
\begin{equation}
\left(\varpi^{X_a}\right)_i = 0.
\end{equation}
In our case, this means we must impose constraints in such a way that respects the invariant subgroup $H_0 =$ SO(2), with unbroken generator $\bar{J}$. But, notice that each of our commutators would necessitate setting
\begin{equation}
\left(\varpi^S\right)_i^{\hphantom{i}j} = 0,
\end{equation}
for different components of $i$ and $j$. In particular, the $\comm{Q}{\bar{P}_i}$ commutator would set the off-diagonal components to zero while the $\comm{D}{\bar{P}_i}$ commutator would set the diagonal components to $0$. Since $\left(\varpi^S\right)_i^{\hphantom{i}j}$ is a tensor under SO(2), we are stuck with two options: either we impose all inverse Higgs constraints, or none.

At first glance, this is not an issue. The more constraints we impose, the simpler the problem should end up being. Let us first impose the following constraint:
\begin{equation}
\label{eq:MFfirstconstraint}
\left(\varpi^S\right)_y^{\hphantom{x}x} = 0 \Rightarrow \tan\pi = \frac{\partial_y \psi^1}{\partial_y \psi^2}.
\end{equation}
The other off-diagonal constraint yields
\begin{equation}
\left(\varpi^S\right)_x^{\hphantom{x}y} = 0 \Rightarrow \tan\pi = -\frac{\partial_x \psi^2}{\partial_x \psi^1}.
\end{equation}
Now let us look at the diagonal components. If we take \autoref{eq:MFfirstconstraint} to be the relation for $\pi$, then 
\begin{equation}
\left(\varpi^S\right)_x^{\hphantom{x}x} = 0 \Rightarrow e^{\left(\Delta + 1\right)\chi} = \frac{\epsilon_{ij} \partial_x \psi^i \partial_y \psi^j}{\sqrt{\delta_{ij}\partial_y \psi^i \partial_y \psi^j}}.
\end{equation}
The final constraint yields
\begin{equation}
\left(\varpi^S\right)_y^{\hphantom{y}y} = 0 \Rightarrow e^{\left(\Delta + 1\right)\chi} = \frac{\delta_{ij}\partial_y \psi^i \partial_y \psi^j}{\sqrt{\delta_{ij}\partial_y \psi^i \partial_y \psi^j}}.
\end{equation}
We will use this final constraint to be our defining relationship for $\chi$. The second and fourth constraints remain and fix extra relations on the field $\psi^i$ and therefore the fields $\sigma^i$. One set of solutions to this constraint, taking $\sigma^i$ to be real, are
\begin{equation}
\label{eq:MFConstraints}
\partial_x\sigma^1 = \partial_y\sigma^2 \qquad \text{and} \qquad \partial_y\sigma^1 = - \partial_x\sigma^2.
\end{equation}
There is a second solution that multiplies the right-hand side of each equation by $-1$, but that solution does not change the remainder of our discussion. One last solution to these constraints are
\begin{equation}
\partial_x\sigma^1 = -1 \qquad \text{and} \qquad \partial_x \sigma^2 = 0.
\end{equation}
The last solution directly leads to nonintegrable fields in the $x$-direction. As a result, we cannot define the notion of modes.

The same conclusion can be drawn from the first solution as well. To see this, let us write these constraints in Fourier space:
\begin{equation}
p_x \tilde{\sigma}^1 = p_y \tilde{\sigma}^2 \qquad \text{and} \qquad p_y \tilde{\sigma}^1 = - p_x \tilde{\sigma}^2.
\end{equation}
Combining these two equations yields
\begin{equation}
\label{eq:MFArtificialConstraint}
\tilde{\sigma}^2 = -\frac{p_y^2}{p_x^2}\tilde{\sigma}^2.
\end{equation}
This is satisfied if $p_x = \pm ip_y$. Therefore, we cannot satisfy the constraints by using candidates with well-defined Fourier modes. Indeed, this can also be seen by looking at particular solutions to \autoref{eq:MFConstraints}, which include
\begin{equation}
\sigma^1 = c_0t + c_1 x + c_2 y + c_3 \qquad \text{and} \qquad \sigma^2 = c_0't -c_2 x + c_1 y + c_3',
\end{equation}
where $c_i$ and $c_i'$ are real constants. As before, these are not integrable solutions.

\subsection{\label{s:MFNoIHC} The no-inverse-Higgs-constraint action}
Since imposing all inverse Higgs constraints yields unphysical dispersion relations, the only option we are left with is to impose no inverse Higgs constraints. This means we will need to work with all the building blocks given in \autoref{eq:MFBuildingBlocks} to construct the following action:
\begin{equation}
    S = \int \mathcal{L}\left(\varpi, \nabla \varpi, \nabla^2 \varpi\right) \det\left(e\right) \dd[3]{x},
\end{equation}
where all spatial indices are contracted so as to respect SO(2) invariance. We are interested in free theories, since fractons only have restricted motion in the free theory, so we will expand to second-order in the fields. This will also allow us to more easily compare theories obtained with and without the inverse Higgs constraint since if they are not equivalent at quadratic order, they are also not equivalent at higher orders. Additionally, we note that, since there are fields that are massive in this theory, we are not justified in restricting the derivative expansion to second-order. However, we will eventually integrate out all of the massive modes in the theory, thereby making the derivative expansion legitimate. At the end of the day, we will want to go up to second-order in derivatives in the massless NG modes, and so we will stop at second-order in the derivative expansion here as well.

It is important to note that this Lagrangian does not include any Wess-Zumino terms. Since we are in a (2 + 1) dimensional spacetime, the Wess-Zumino terms will be exact 4-forms built by wedging the Maurer-Cartan coefficients together in an invariant fashion. Since the Maurer-Cartan form is a one-form, the Wess-Zumino terms will be of order four in the Maurer-Cartan coefficients, and therefore the fields. Since we truncate the theory at second-order in the fields, they do not contribute in this case. For specific examples illustrating the construction of Wess-Zumino terms and why they would not contribute in our case, see \cite{Nicolis:2013lma}.

The computation of a theory to second-order in fields and derivatives of the Maurer-Cartan coefficients is a computationally tedious task, but can be accomplished using tensor index packages such as \verb!xTensor!. The idea is to create a list of every Maurer-Cartan coefficient and then generate a list of all possible combinations of Maurer-Cartan coefficients. After that, \verb!xTensor! can take these combinations and compute all possible index contractions. The details of this computation and the subsequent expansion to second order in fields and derivatives are left to the supplemental material \cite{SuppMat}. For now, we state the final result:
\begin{equation}
\label{eq:MFLagrangian}
\begin{aligned}
\mathcal{L}\left(\sigma^i, \pi, \chi\right)\det\left(e\right) &= \frac{1}{2}\partial_t \sigma_i \partial_t \sigma^i - \kappa_1 \partial_i \sigma^i \partial_j \sigma^j - \kappa_2 \partial_i \sigma^j \partial^i \sigma_j \\
&\hspace{0.5cm} - \frac{1}{2} m_{\pi}^2 \pi^2 - \frac{1}{2} m_\chi^2 \chi^2 + g \epsilon^{ij} \pi \partial_i \sigma_j \\
&\hspace{0.5cm} + \frac{1}{2}\left(\left(\Delta + \frac{1}{2}\right)m_\pi^2 - 8\left(\Delta + 1\right)\kappa_1 - 3 g\right)\chi \partial_i \sigma^i.
\end{aligned}
\end{equation}
The first line in the Lagrangian contains the kinetic terms while the second line contains the mass and cross terms, where we note that the masses are not the physical masses of the fields, rather just the coefficients of the quadratic term. The last term is also a cross term, but its coefficient depends on the remaining parameters in the given way. Note that we do not include any kinetic terms of the fields $\pi$ and $\chi$. This is due to the fact we are mostly concerned with expanding to second-order in derivatives of the massless field $\sigma^i$. Indeed, we will shortly see that $\pi$ and $\chi$ are related to the derivative of $\sigma^i$ upon use of the equations of motion to integrate these fields out. Therefore, we consider a term like $\pi \partial_i \sigma_j$ to be second-order in derivatives of $\sigma^i$, while a kinetic term such as $\left(\partial_t\pi\right)^2$ is fourth order in derivatives of $\sigma^i$.\footnote{This argument requires the masses of $\pi$ and $\chi$ to be nonzero. However, the masses are free parameters from the perspective of the coset construction, so there are a subset of theories for which $\pi$ and $\chi$ are massless and we will need to expand to second-order in those fields anyway. While this is not important for our purposes, this full Lagrangian is contained at the end of the code in the supplemental material \cite{SuppMat}.} This allows us to simplify the Lagrangian without affecting the remainder of the calculation and results.

Now let us look at the equation of motion for $\pi$ and $\chi$ in order to integrate out these massive fields, noting that while the masses $m_\pi$ and $m_\chi$ are not themselves the physical masses, they do parametrize the physical masses: 
\begin{equation}
    \begin{aligned}
        \pi = \frac{g}{m_{\pi}^2}\epsilon^{ij} \partial_i \sigma_j \qquad \text{and} \qquad \chi = \frac{g'}{m_\chi^2}\partial_i\sigma^i,
    \end{aligned}
\end{equation}
where $g'$ is the constrained coefficient of $\chi\partial_i\sigma^i$ in the Lagrangian. Plugging this back into the Lagrangian, we obtain an effective field theory solely for $\sigma^i$:
\begin{equation}
\label{eq:sigmaiEFT}
    \mathcal{L}\left(\sigma^i\right)\det\left(e\right) = \frac{1}{2}\partial_t \sigma_i \partial_t \sigma^i - \tilde{\kappa}_1 \partial_i \sigma^i \partial_j \sigma^j - \kappa_2 \partial_i \sigma^j \partial^i \sigma_j + \tilde{g}^2 \epsilon^{ij} \epsilon^{kl} \partial_i \sigma_j \partial_k \sigma_l,
\end{equation}
where 
\begin{equation}
\tilde{\kappa}_1 = \kappa_1 - \frac{\left(g'\right)^2}{2m_\chi^2} \qquad \text{and} \qquad \tilde{g}^2 = \frac{g^2}{2m_\pi^2}.
\end{equation}
As an important sanity check that everything we have done to this point is correct, we can compare this result to \cite[2.71]{Argurio:2021opr}, which we quoted in \autoref{eq:ArgurioMF}. As a nontrivial check, we included \textit{some} terms that are fourth order in derivatives of $\sigma^i$ and found that by fine tuning constants such as $\tilde{\kappa}_1$, $\kappa_2$, and $\tilde{g}^2$ to $0$, we obtain the same Lagrangian. For our present purposes, we stick with the presented Lagrangian up to second-order in derivatives.

To study the physical consequences of this effective Lagrangian, we should study the dispersion relations of the particles associated to $\sigma^i$. To do this, we first express the Lagrangian as $\sigma^i K_{ij} \sigma^j$, where $K_{ij}$ is the kinetic matrix. In momentum space,
\begin{equation}
    \tilde{K}_{ij} = \frac{1}{2}\delta_{ij}\left(\omega^2 - 2\kappa_2 p_k p^k\right) - \tilde{\kappa}_1 p_i p_j + \tilde{g}^2\epsilon_{ik} \epsilon_{jl} p^k p^l.
\end{equation}
The dispersion relations can be found by setting the determinant of this matrix to $0$ and solving for $\omega$, which yields two massless modes
\begin{equation}
    \begin{aligned}
        \omega_1^2 &= 2\left(\kappa_2 - \tilde{g}^2\right)\left(p_x^2 + p_y^2\right),\\
        \omega_2^2 &= 2\left(\tilde{\kappa}_1 + \kappa_2\right)\left(p_x^2 + p_y^2\right).
    \end{aligned}
\end{equation}
Notice that if $\tilde{g}^2 = \kappa_2$, we have $\omega_1 = 0$. Therefore, to second-order in the derivative expansion, one mode is a pure fractonic mode. There is a symmetry explanation for the vanishing of $\omega_1^2$. Upon setting $\tilde{g} = \kappa_2$, the Lagrangian, \autoref{eq:sigmaiEFT}, is invariant under the transformation
\begin{equation}
    \sigma^i \rightarrow \sigma'^{\:i} = \sigma^i + \epsilon^{ij}\partial_j f\left(x, y\right),
\end{equation}
for some generic function $f\left(x, y\right)$. The fact that such emergent symmetries under shifts by nonconstant functions are associated to fractonic behavior in the dispersion relations is discussed in \cite{Pretko:2018jbi}.

It is interesting to note that this symmetry also exists at the level of the Lagrangian prior to integrating out the massive modes, \autoref{eq:MFLagrangian}, when setting $g = \kappa_2 = 0$. This is easiest to see since the Lagrangian solely becomes a function of $\pi$, $\chi$, and $\partial_i \sigma^i$. The inclusion of terms that treat $\pi$ and $\chi$ at the same order in the derivative expansion of $\sigma^i$ does not affect this result. That Lagrangian, without any fine tuning, is
\begin{equation}
\label{eq:FullerMFLag}
\begin{aligned}
\mathcal{L}\left(\sigma^i, \pi, \chi\right)\det\left(e\right) &= \frac{1}{2}\partial_t \sigma_i \partial_t \sigma^i + \frac{1}{2} \left(\partial_t \pi\right)^2 + \frac{1}{2}\left(\partial_t \chi\right)^2 - \frac{1}{2} m_\pi^2 \pi^2 - \frac{1}{2} m_\chi^2 \chi^2 \\
&\hspace{0.5cm} - \kappa_1' \partial_i \pi \partial^i \pi - \kappa_2' \partial_i \chi \partial^i \chi - \kappa_1 \partial_i \sigma^i \partial_j \sigma^j - \kappa_2 \partial_i \sigma^j \partial^i \sigma_j \\
&\hspace{0.5cm} + \kappa_3' \chi \partial_t \pi + \kappa_4' \partial_t \chi \partial_t \pi + \kappa_5' \partial_i \chi \partial^i \pi \\
&\hspace{0.5cm} + \kappa_6' \partial_t \chi \partial_i \sigma^i + \kappa_7' \partial_t \pi \partial_i \sigma^i + g \epsilon^{ij} \pi \partial_i \sigma_j + g_7 \epsilon^{ij} \partial_i \pi \partial_t \sigma_j \\
&\hspace{0.5cm} + \frac{1}{2}\left(\left(\Delta + \frac{1}{2}\right) m_\pi^2 - 8\left(\Delta + 1\right)\kappa_1 - 3g\right) \chi \partial_i \sigma^i,
\end{aligned}
\end{equation}
where any terms that are present here that were not in \autoref{eq:MFLagrangian} have a coefficient $\kappa_i'$. This is still only a function of $\pi$, $\chi$, $\partial_i\sigma^i$, and $\partial_t \sigma^i$ when $g = \kappa_2 = 0$. As a result, it still has the same enhanced shift symmetry on $\sigma^i$.

\section{\label{s:HSF} The Helical Superfluid}
Turning our attention to the second symmetry-breaking pattern given by \autoref{HSFgen}, the nontrivial commutators of the resulting Lie algebra are: 
\begin{equation}
\label{eq:HSFLieAlg}
\begin{aligned}
\comm{Q}{S_i} &= -i\epsilon_i^{\hphantom{i}j} S_j, \qquad & \qquad \comm{D}{\bar{P}_\mu} &= i \left(\bar{P}_\mu - k\delta_\mu^{\hphantom{\mu}1}Q\right), \\
\comm{S_i}{\bar{P}_x} &= ik\epsilon_i^{\hphantom{i}j} S_j, & \comm{J}{\bar{P}_i} &= -i\epsilon_i^{\hphantom{i}j}\left(\bar{P}_j - k\delta_j^{\hphantom{j}1}Q\right), \\
\comm{S_i}{D} &= i\Delta S_i. & &
\end{aligned}
\end{equation}
Recall that by \autoref{HSFgen}, $H_0$ is trivial, so all broken generators and unbroken translations are multiplets of $H_0$ and we are able to perform the coset construction. Once again following \cite{Klein:2017npd}, we adopt the coset parametrization that contains the most number of exponentials. Notice that we again use the fact that $\comm{S_i}{S_j}=0$, so we can combine the $S_i$ components into one exponential without loss of generality:
\begin{equation}
\label{eq:OGHSFparametrization}
U\left(x, \pi, \sigma^i, \chi, \theta\right) = e^{ix^\mu \bar{P}_\mu} e^{i\pi Q} e^{i\sigma^i S_i} e^{i\chi D} e^{i\theta J}.
\end{equation}
For computational simplicity, we can take $\psi \equiv \pi + kx$ and write
\begin{equation}
\label{eq:HSFparametrization}
U\left(x, \psi, \sigma^i, \chi, \theta\right) = e^{ix^\mu P_\mu} e^{i\psi Q} e^{i\sigma^i S_i} e^{i\chi D} e^{i\theta J}.
\end{equation}
The same subtleties that we must make sure to expand to second-order in $\pi$ instead of $\psi$ as well as write the Maurer-Cartan form using $\bar{P}_\mu$ instead of $P_\mu$, remain, but these two parametrizations are checked to be equivalent.

As before, we will use the Maurer-Cartan form to give us objects covariant under the full symmetry group with which to build an invariant action. After immediate simplifications using commutators that are zero,
\begin{equation}
\begin{aligned}
U^{-1}\partial_\mu U &= i\Big(e^{-i\theta J} e^{-i\chi D} P_\mu e^{i\chi D} e^{i\theta J} + e^{-i\chi D} e^{-i\sigma^i S_i} \partial_\mu \psi Q e^{i\sigma^i S_i} e^{i\chi D} \\
&\hspace{1cm} + e^{-i\chi D} \partial_\mu \sigma^i S_i e^{i\chi D} + \partial_\mu \chi D + \partial_\mu \theta J\Big).
\end{aligned}
\end{equation}
Using the results in \autoref{eq:HSFMCformhelp}, we can write the final Maurer-Cartan form with $\bar{P}_\mu$ as
\begin{equation}
\begin{aligned}
U^{-1}\partial_\mu U &= i\Big(e^{\chi} R_\mu^{\hphantom{\mu}\nu}\left(-\theta\right)\bar{P}_\nu + \left(\partial_\mu \psi - k e^{\chi} R_\mu^{\hphantom{\mu}1}\left(-\theta\right)\right)Q \\
&\hspace{1cm} + e^{-\Delta \chi}\left(\partial_\mu \sigma^i - \epsilon^i_{\hphantom{i}j}\sigma^j \partial_\mu \psi\right)S_i + \partial_\mu \chi D + \partial_\mu \theta J\Big).
\end{aligned}
\end{equation}
Referencing \autoref{eq:GenericMCForm}, the coset tetrad and Maurer-Cartan coefficients are
\begin{equation}
\label{eq:HSFMCcoefs}
\begin{aligned}
e_\mu^{\hphantom{\mu}\nu} &= e^\chi R_\mu^{\hphantom{\mu}\nu}\left(-\theta\right), \qquad & \qquad  \left(\varpi^D\right)_\mu &= e^{-\chi}R_{\mu}^{\hphantom{\mu}\nu}\left(\theta\right)\partial_\nu \chi, \\
\left(\varpi^Q\right)_\mu &= e^{-\chi} R_{\mu}^{\hphantom{\mu}\nu}\left(\theta\right)\partial_\nu \psi - k\delta_\mu^{\hphantom{\mu}1}, & \left(\varpi^J\right)_\mu &= e^{-\chi} R_{\mu}^{\hphantom{\mu}\nu}\left(\theta\right)\partial_\nu \theta, \\
\left(\varpi^S\right)_\mu^{\hphantom{\mu}i} &= e^{-\left(\Delta + 1\right)\chi}\left(\partial_\mu \sigma^i - \epsilon^i_{\hphantom{i}j}\sigma^j \partial_\mu \psi\right). & &
\end{aligned}
\end{equation}
Note that the Maurer-Cartan forms are first order in the fields. One can check that setting $\pi = \sigma^i = \chi = \theta = 0$ sets all the Maurer-Cartan forms to $0$. Additionally, we checked that all components of these Maurer-Cartan coefficients are invariant under the transformations in \hyperref[app:TransformationLaws]{Appendix D}, as expected by the fact that $H_0$ is trivial. Lastly, the invariant integration measure and covariant derivative from \autoref{eq:GeneralCovD} are
\begin{equation}
    \det\left(e\right)\dd[3]{x} = e^{3\chi}\dd[3]{x} \qquad \text{and} \qquad \nabla_\mu = e^{-\chi} R_\mu^{\hphantom{\mu}\nu}\left(\theta\right)\partial_\nu.
\end{equation}

Analogous to the metafluid, we will employ two distinct approaches to get rid of the massive modes and obtain an EFT solely for massless modes: one that imposes inverse Higgs constraints and one that does not. As before, based on the Lie algebra, \autoref{eq:HSFLieAlg}, and the arguments found in \hyperref[app:IHCReview]{Appendix B}, there will be four massive terms.\footnote{Given that we have five broken generators, this leaves \textit{a priori} only one massless field. Since there is only one massless field, by Goldstone's theorem we will have exactly one NG mode.} Through this, we will see that the theories obtained by imposing inverse Higgs constraints are a subset of the theories obtained by writing down a more general massive theory and integrating out the massive modes.

\subsection{The inverse-Higgs-constraint-imposed action}

In order to impose inverse Higgs constraints, we must first look at the commutators involving two broken generators and unbroken translations: 
\begin{equation}
\begin{aligned}
\comm{D}{\bar{P}_x} &= i \left(\bar{P}_x - kQ\right), \qquad & \qquad 
\comm{S_1}{\bar{P}_x} &= ik S_2,\\
\comm{S_2}{\bar{P}_x} &= -ik S_1, \qquad & \qquad \comm{J}{\bar{P}_y} &= -i\left(\bar{P}_x - kQ\right).
\end{aligned}
\end{equation}
Unlike the metafluid we do not have unbroken rotational symmetry, so there is a greater flexibility in the number of inverse Higgs constraints that can be imposed. In other words, because $H_0$ is trivial, any number of constraints can be imposed and the theory would still be invariant under the full symmetry group, $G$.

The commutation relations suggest the existence of four inverse Higgs constraints that can be applied. However, imposing two of these constraints simultaneously introduces an ambiguity. Specifically, the commutators
\begin{equation}
    \comm{S_1}{\bar{P}_x} = i k S_2 \qquad \text{and} \qquad \comm{S_2}{\bar{P}_x} = -i k S_1
\end{equation}
imply that we can either express $\sigma^1$ in terms of derivatives of $\sigma^2$ or $\sigma^2$ in terms of derivatives of $\sigma^1$. Imposing both simultaneously would lead to an NG candidate expressed in terms of its own derivative. 

While we will explore the situation involving the imposition of all four inverse Higgs constraints later in this section, we chose to impose three inverse Higgs constraints at the moment to avoid any potential ambiguities. In particular, we will exclude one of the two ``troublesome" commutators in alignment with established practices in the literature, such as \cite[Appendix C]{Bonifacio2019}. With these criteria in mind, we set the following three Maurer-Cartan coefficients to zero:
\begin{equation}
\left(\varpi^Q\right)_2 = \left(\varpi^Q\right)_1 = \left(\varpi^S\right)_1^{\hphantom{1}1} = 0.
\end{equation}
Doing so, as expected of imposing inverse Higgs constraints, establishes the following relationships between the fields:
\begin{equation}
\label{HelicalFluid1stIHCResult}
\tan\theta = -\frac{\partial_y\psi}{\partial_x\psi},
\end{equation}
\begin{equation}
\label{eq:chitopsi}
ke^{\chi} = \sqrt{\delta^{ij}\partial_i\psi \partial_j \psi},
\end{equation}
\begin{equation}
\sigma^2 = \frac{\delta^{ij}\partial_i\psi\partial_j\sigma^1}{\delta^{ij}\partial_i\psi\partial_j\psi}.
\label{HelicalFluid3rdIHCResult}
\end{equation}
We use the Maurer-Cartan coefficients as building blocks of our Lagrangian and write down all the possible combinations of terms that are up to second-order in fields and derivatives. In contrast to the metafluid scenario, we are not constrained by the need to carefully manage contractions to preserve any SO(2) symmetry or similar constraints, as all generators are broken. In other words, the action is of the form
\begin{equation}
S = \int \mathcal{L}\left(\varpi, \nabla \varpi, \nabla^2\varpi\right)\det\left(e\right)\dd[3]{x},
\end{equation}
where we go up to second-order in Maurer-Cartan forms. After that, we will integrate out the remaining massive mode. Analogous to the metafluid case, because we keep the theory to second-order in the fields and the Wess-Zumino terms are fourth order in the Maurer-Cartan coefficients and fields, such terms are excluded from consideration.

As with the metafluid, we will write down the resulting Lagrangian up to second-order in the massless fields and treat the massive fields as being the same order as a derivative of the massless field. This simplification is justified as all massive fields, upon integration, are indeed derivatives of the massless fields and so their exclusion yields a simpler Lagrangian and does not alter any of our remaining results. We insert the constraints given by \autoref{HelicalFluid1stIHCResult}-\autoref{HelicalFluid3rdIHCResult} and write down every unique combination of linear and quadratic Maurer-Cartan forms yielding
\begin{equation} 
\label{eq:HSFpisigmaLag}
\mathcal{L}\left(\pi, \sigma\right)\det\left(e\right) = \frac{1}{2}\left(\partial_t\pi\right)^2 - \kappa \left(\partial_y\pi\right)^2 + g_1\sigma \partial_t \pi + g_2 \sigma \partial_x \pi - \frac{1}{2}m_\sigma^2 \sigma^2.
\end{equation}
This result did not require any use of \verb!xTensor! as the Maurer-Cartan coefficients all simplified greatly. For notational simplicity, we let $\sigma^1 = \sigma$.

As stated before, we now integrate out the massive $\sigma$ field and observe that it is expressed as a derivative of $\pi$. Up to first order in derivatives, the equation of motion is:
\begin{equation}
\sigma = \frac{1}{m_\sigma^2}\left(g_1 \partial_t \pi + g_2 \partial_x \pi\right).
\end{equation}
Substituting this expression back into the Lagrangian, and adhering to our hypothesis of retaining terms up to second-order in the fields and their derivatives, 
\begin{equation}
\label{hsfEFTIHC}
\mathcal{L}\left(\pi\right)\det\left(e\right) = \frac{1}{2}\left(\partial_t\pi\right)^2 - \tilde{g}^2\left(\partial_x\pi\right)^2 - \tilde{\kappa} \left(\partial_y\pi\right)^2 - 2\gamma \tilde{g}^2\partial_t \pi \partial_x\pi,
\end{equation}
where we recanonically normalized the $\pi$ field and defined
\begin{equation}
\tilde{g}^2 = \frac{g_2^2}{2m_\sigma^2\left(1 + g_1^2/m_\sigma^2\right)}, \qquad \tilde{\kappa} = \frac{\kappa}{1 + g_1^2/m_\sigma^2}, \qquad \text{and} \qquad \gamma = -\frac{g_1}{g_2}.
\end{equation}

Taking the kinetic matrix, as before, and setting the determinant to $0$, we find the following dispersion relations:
\begin{equation}
\omega = 2\tilde{g}^2\gamma p_x \pm \sqrt{2\tilde{g}^2\left(1 + 2\tilde{g}^2\gamma^2\right)p_x^2 + 2\tilde{\kappa} p_y^2}.
\end{equation}
Notice that we end up with double dispersion relations despite the presence of only a single real scalar field. Mathematically, this can be attributed to the competition between quadratic and linear time derivative terms. In fact, by setting $\gamma$ to $0$, we recover more standard dispersion relations of anisotropic phonons, $\omega^2 = 2\tilde{g}^2 p_x^2 + 2\tilde{\kappa} p_y^2$. Physically, nonrelativistic systems with broken spacetime symmetries, such as the one we are considering, have been shown to exhibit more complex dispersion relations than those found in Poincaré-invariant quantum field theories (QFTs) \cite{Watanabe_2014}. For similar discussions for a field theory that features several dispersion relations associated to one single real scalar field in the context of phonons, see \cite{Musso_2020}.

An interesting feature of the dispersion relations emerges when $\tilde{\kappa} = 0$. This condition eliminates the $y$-dependent derivatives in the action, resulting in an emergent shift symmetry $\pi \rightarrow \pi + g(y)$, where $g(y)$ is a totally generic function of $y$. In the dispersion relations, the momentum term in the $y$-direction vanishes, such that $\omega \propto p_x$. Consequently, motion is restricted from propagating in the $y$-direction, exhibiting fractonic (lineon) behavior. Similarly, by setting $\tilde{g} = 0$, all $p_x$ dependence is eliminated in the dispersion relation, resulting in an emergent shift symmetry in the Lagrangian of the form $\pi \rightarrow \pi + h(x)$, where $h(x)$ is a completely arbitrary function of $x$. This leads to dispersion relations $\omega \propto p_y$, and the resulting fractonic (lineon) motion is restricted from propagating in the $x$-direction.

Having carefully examined the action and effective field theory under the imposition of three inverse Higgs constraints, we now consider the case of all four inverse Higgs constraints. Imposing the fourth inverse Higgs constraint to zero, i.e. $\left(\varpi^S\right)_1^{\hphantom{1}2} = 0$, implies the following relationship between the fields:
\begin{equation} \label{hsfconstraint}
\sigma^1 = -\frac{\delta^{ij}\partial_i\psi\partial_j\sigma^2}{\delta^{ij}\partial_i\psi\partial_j\psi}.
\end{equation}
Then, the other constraints \autoref{HelicalFluid1stIHCResult}-\autoref{HelicalFluid3rdIHCResult} yield
\begin{equation}
\label{eq:HSFArtificialConstraint}
    \sigma^1 = - \frac{1}{k^2}\partial_1^2 \sigma^1. 
\end{equation}
Notice that \autoref{eq:HSFArtificialConstraint}  mixes the derivative order, thereby invalidating the truncated derivative expansion. In particular, it is unclear how to make the derivative expansion consistent because any order of $\sigma^1$ can be represented as higher-order derivatives or reduced-order. In addition, the coset construction relies on our ability to define a derivative expansion. A second issue appears when we Fourier-transform the constraint:
\begin{equation}
    \tilde\sigma^1(\omega, p) = \frac{p_x^2}{k^2}\tilde\sigma^1(\omega, p) \quad \Leftrightarrow \quad p_x=\pm k.
\end{equation}
This constrains the solutions for $\sigma^1$ to be plane waves propagating along the $x$-direction with wave vector $k$, where we recall that $k$ is a single constant appearing in the Lie algebra. The imposition of the fourth inverse Higgs constraint results in an EFT that is not physically equivalent to the EFT derived through imposing three inverse Higgs constraints and integrating out of the massive mode, $\sigma^1$ [which was relabeled to $\sigma$ in \autoref{eq:HSFpisigmaLag}], since in the latter case, there was no mode with a constrained wave-vector. As a result, imposing the last inverse Higgs constraint seems like an artificial constraint rather than a physical one. 

\subsection{The no-inverse-Higgs-constraint action}
Since there are conceptual issues involving the use of four inverse Higgs constraints, it is worthwhile to wonder whether imposing three inverse Higgs constraints would be consistent with integrating out massive modes. Let us return to the Maurer-Cartan coefficient in \autoref{eq:HSFMCcoefs} and write every combination of $\varpi$, $\nabla\varpi$, and $\nabla^2\varpi$ possible and expand up to second-order in fields and derivatives. The computation is best handled using \verb!xTensor! and is left to the Supplemental Material \cite{SuppMat}. The resulting Lagrangian is stated below:
\begin{equation}
\label{eq:HSFMassiveEFT}
\begin{aligned}
\mathcal{L}\det\left(e\right) &= \frac{1}{2}\left(\partial_t\pi\right)^2 - \kappa_1 \left(\partial_x\pi\right)^2 - \frac{m_{\theta}^2}{2k^2}\left(\partial_y\pi\right)^2 - \kappa_2 \partial_t \pi \partial_x\pi - \kappa_3 \partial_t \pi \partial_y \pi - \kappa_4 \partial_x \pi \partial_y \pi \\
&\hspace{0.5cm} - \frac{1}{2} m_{\sigma^1}^2 \left(\sigma^1\right)^2 - \frac{1}{2} m_{\sigma^2}^2 \left(\sigma^2\right)^2 - \frac{1}{2} m_{\chi}^2 \chi^2 - \frac{1}{2}m_{\theta}^2\theta^2\\
&\hspace{0.5cm} + \kappa_5 \sigma^1 \sigma^2 + \kappa_6 \sigma^1 \chi + k \kappa_7 \sigma^1 \theta + \kappa_8 \sigma^2 \chi + k \kappa_9 \sigma^2 \theta + k\kappa_{10} \chi \theta\\
&\hspace{0.5cm} + \kappa_{11} \sigma^1 \partial_t \pi + \kappa_{12} \sigma^2 \partial_t \pi + \kappa_{13} \chi \partial_t \pi + k\kappa_3 \theta \partial_t \pi \\
&\hspace{0.5cm} + \kappa_{14} \sigma^1 \partial_x \pi + \kappa_{15} \sigma^2 \partial_x \pi + 2\left(\kappa_{16} + k \kappa_1\right)\chi \partial_x \pi + \frac{1}{2}\left(3k \kappa_4 + \kappa_{10}\right)\theta \partial_x \pi \\
&\hspace{0.5cm} - \kappa_7 \sigma^1 \partial_y \pi - \kappa_9 \sigma^2 \partial_y \pi - \kappa_{10} \chi \partial_y \pi + \left(\kappa_{16} + \frac{m_{\theta}^2}{k}\right)\theta \partial_y \pi,
\end{aligned}
\end{equation}
where $\kappa_i$ are free parameters and $m_i$ are the coefficients of the quadratic terms, which parametrize the physical masses but are not exactly the physical masses.\footnote{Note that before relabeling any coefficients as we have done here, the mass terms of $\sigma^1$, $\sigma^2$, and $\theta$ are all proportional to $k^2$, while the mass term of $\chi$ is a quadratic polynomial in $k$, where $k$ is the Lie algebra parameter. It would be interesting to further explore the relationship between the mass gap and $k$ which arises from the homogeneous breaking of spatial translations, similar to how it was done in \cite{Nicolis:2013sga} for the chemical potential, $\mu$, arising from the homogeneous breaking of time translations.} We reemphasize that although the heavy fields $\sigma$, $\chi$, and $\theta$ do not contain any kinetic terms, they are not auxiliary fields since the kinetic terms would appear if we continued the derivative expansion. 

Prior to integrating out the masses, let us investigate the relationship between the inverse Higgs constraints in \autoref{HelicalFluid1stIHCResult}-\autoref{HelicalFluid3rdIHCResult} and the equations of motion for the massive modes. To first order in the fields $\pi$ and $\chi$ the constraint equations \autoref{HelicalFluid1stIHCResult}-\autoref{HelicalFluid3rdIHCResult} yield:
\begin{equation}
\label{eq:OldConstraintsEOM}
\theta = -\frac{1}{k}\partial_y \pi \qquad \text{and} \qquad \chi = \frac{1}{k}\partial_x \pi.
\end{equation}
If we instead impose the equations of motion for $\theta$, $\chi$, $\sigma^1$, and $\sigma^2$ using the Lagrangian \autoref{eq:HSFMassiveEFT}, we find 
\begin{equation}\label{eq:eomthetaHSF}
\theta = -\frac{1}{k}\partial_y \pi + \alpha_1 \partial_x \pi + \alpha_2 \partial_t \pi,
\end{equation}
and 
\begin{equation} \label{eq:eomchiHSF}
\chi = \alpha_3 \partial_x \pi + \alpha_4 \partial_t \pi.
\end{equation} 
where $\alpha_1$, $\alpha_2$, $\alpha_3$, and $\alpha_4$ are nontrivial combinations of all the coupling constants that appear in \autoref{eq:HSFMassiveEFT} whose exact expression is not particularly illuminating but can be found in the supplemental code \cite{SuppMat}. Here, we see that the equations of motion are not equivalent to the inverse Higgs constraints. At this order, it is the $\partial_t\pi$ term in \autoref{eq:eomchiHSF} that differentiates it from the form of $\chi$ that is found via the inverse Higgs constraint. Similarly, there are additional $\partial_x \pi$ and $\partial_t \pi$ terms in \autoref{eq:eomthetaHSF} compared to the form of $\theta$ found via the inverse Higgs constraints. Let us mention that out of the complicated expressions of the $\alpha_i$ in terms of the coupling constants, it is possible to fine-tune \autoref{eq:eomthetaHSF} and \autoref{eq:eomchiHSF} to \autoref{eq:OldConstraintsEOM}. If such a fine tuning is selected, we would recover the case of the previous subsection, since the additional Maurer-Cartan forms included in the derivation of this EFT would be set to zero as was done immediately using the Inverse Higgs Constraints.

Although at this order the inverse Higgs constraints do not produce the same equation for the massive fields as directly integrating the mass out, it should still be verified explicitly whether the resulting effective field theory is equivalent to the inverse-Higgs-constraints-imposed EFT \autoref{hsfEFTIHC} modulo field redefinitions. What we find, however, is that additional $\partial_t \pi$ terms introduce new contributions to the no-inverse-Higgs-constraints effective Lagrangian:
\begin{equation}
\label{eq:HSFFinalLag}
\mathcal{L}\left(\pi\right)\det e = \left(\partial_t\pi\right)^2 + \tilde{\kappa}_1 \left(\partial_x \pi\right)^2 + \tilde{\kappa}_2 \left(\partial_y \pi\right)^2 + \tilde{\kappa}_3 \partial_t \pi \partial_x \pi + \tilde{\kappa}_4 \partial_t \pi \partial_y \pi + \tilde{\kappa}_5 \partial_x \pi \partial_y \pi,
\end{equation}
where all of the tilded quantities are rewritings of a nontrivial combination of the coupling constants in the previous Lagrangian, \autoref{eq:HSFMassiveEFT}, noting that all the $\tilde{\kappa}_i$ are independent. The important point here is that there are two new operators that appear in this Lagrangian that did not appear before, $\partial_x \pi \partial_y \pi$ and $\partial_t\pi \partial_y \pi$. The presence of these operators makes this a qualitatively different theory compared to the one where three inverse Higgs constraints are imposed in which these operators are absent.

To conclude this section, we compute the dispersion relations to be:
\begin{equation}
    \omega = 
   \frac{1}{2} \left(\tilde{\kappa}_3 p_x + \tilde{\kappa}_4 p_y \pm 
      \sqrt{(\tilde{\kappa}_3 p_x + \tilde{\kappa}_4 p_y)^2 - 
       4 (\tilde{\kappa}_1 p_x^2 +  \tilde{\kappa}_2 p_y^2 + 
           \tilde{\kappa}_5 p_x p_y)}\right).
\end{equation}
It can be seen that fractonic behavior can be obtained in two different ways. First, we can set $\tilde{\kappa}_2$ and $\tilde{\kappa}_4$ to $0$, which produces a fractonic dispersion relation in which the mode cannot propagate strictly along $y$. This corresponds to a shift symmetry in the Lagrangian of the form $\pi \rightarrow \pi + f\left(y\right)$ for a generic function $f\left(y\right)$. Equivalently, we could set $\tilde{\kappa}_1$ and $\tilde{\kappa}_3$ to $0$ and obtain fractonic behavior in which the mode cannot propagate exactly along $x$ with an associated shift symmetry $\pi \rightarrow \pi + f\left(x\right)$ in the Lagrangian. Alternatively, we could additionally set $\tilde{\kappa}_5$ to zero in both cases, which would yield pure lineon behavior since the dispersion relations would be linear in $p_x$ in the first case and $p_y$ in the second. One difference between the effective theories with and without imposing inverse Higgs constraints at the level of the dispersion relations is that, here, we need to set two parameters to $0$ to obtain fractonic behavior, whereas in the case where we imposed inverse Higgs constraints, we only needed to set one parameter to $0$, once again indicating that the inverse Higgs constraints play the role of a fine tuning that sets certain parameters to $0$.
\newpage
\section{\label{s:Discussion} Discussion and Conclusion}
In this paper, we used the coset construction to study two different spontaneous symmetry-breaking patterns. The first symmetry-breaking pattern homogeneously broke the group of spatial translations, rotations, dilations, U(1), and complex shifts to a diagonal combination of rotations and U(1) and a diagonal combination of translations and shifts. The associated theory was called the metafluid. The second symmetry-breaking pattern homogeneously broke the same group to a diagonal combination of translations and U(1). The associated theory was called the helical superfluid. 

The first main result of the paper is that imposing inverse Higgs constraints when translations are broken homogeneously can lead to a loss of generality. For the metafluid, we found that imposing all inverse Higgs constraints resulted in theories which yielded solutions that are divergent over spatial infinity and could not be used to properly define Fourier modes. On the other hand, when we avoided imposing inverse Higgs constraints and instead wrote down a generic theory invariant under the symmetry group and integrated out the massive modes by hand, we found a theory whose solutions have proper Fourier transforms, yielding dispersion relations with real frequencies. In the case of the helical superfluid, when we imposed three out of the four possible inverse Higgs constraints, this inequivalence manifested as two Lagrangians with a different number of independent operators. Indeed, the theory with inverse Higgs constraints imposed, \autoref{hsfEFTIHC}, can be viewed as taking a more generic theory without any inverse Higgs constraints imposed, \autoref{eq:HSFFinalLag}, and fine tuning certain parameters to zero. When we imposed all four inverse Higgs constraints, we found a nonlocal constraint that relates the field to derivatives of that same field, which seems \textit{a priori} artificial and mixes up orders in the derivative expansion. 

It is important to note that the homogeneous breaking of translations in our examples is a reason why some of the eventual inverse Higgs constraints led to unphysical theories. A potential inverse Higgs constraint exists when the commutator of an unbroken translation generator with a broken generator includes another broken generator. In the case where translations are broken homogeneously, we must use the unbroken translation generator $\bar{P}_\mu$, which is a linear combination of $P_\mu$ and some other internal broken generator $X$. Consequently, if $X$ does not commute with another broken generator, $Y$, then $\comm{\bar{P}}{Y}$ will not be zero even if $\comm{P}{Y}$ is, which introduces additional potential inverse Higgs constraints. To clarify, at the level of the algebra, homogeneous breaking of translations causes a mixing between internal symmetries and spacetime symmetries, meaning even internal symmetries can now contribute to inverse Higgs constraints. This introduces two potential ambiguities. First, the various commutators and resulting inverse Higgs constraints are more likely to be interdependent, as occurred in \autoref{eq:HSFLieAlg} with the shift generators. Second, the presence of more inverse Higgs constraints than massive fields may result in an overconstrained system, such as \autoref{eq:MFArtificialConstraint}, or constraints that break the derivative expansion, such as \autoref{eq:HSFArtificialConstraint}. We recall that in the metafluid, this was unavoidable as the presence of unbroken rotations meant we either had to impose all or none of the inverse Higgs constraints, so it was impossible to pick a subset of inverse Higgs constraints that did not overly constrain the system. Suppose we instead look at the algebra involving just $P_\mu$ instead of $\bar{P}_\mu$, \autoref{eq:FullLieAlg}, thus going back to the case where translations are not broken. Then, the commutators that originally led to the inverse Higgs constraints no longer contain a broken generator on the right-hand side. Since there would no longer be any inverse Higgs constraint to impose in this case, it trivializes the above discussion.

To this point, we have attempted to answer the question of whether imposing the inverse Higgs constraint leads to a loss of generality. Another perspective on this question is to ask whether constructing a theory directly from the transformation laws of the massless NG candidates depends on how one eliminates the other NG candidates. This is explored in detail in \cite{Klein:2017npd}. In particular, they assume a generic transformation of the coordinates, $x^\mu$, massless NG candidates $\phi^a$, and massive NG candidates, $\xi^m$, such that the covariant derivative is an irreducible representation of the unbroken subgroup without translations, $H_0$, and that the Maurer-Cartan coefficients $\varpi^a_\mu$ are independent of $\partial_\mu \xi^m$ and transform covariantly under the symmetry group $G$, so that if an inverse Higgs constraint can be imposed, we could set $\varpi^a_\mu = 0$\footnote{In the paper, the authors refer to $\phi^a$ as essential and $\xi^m$ as inessential NG candidates.}. With this, they prove that the transformation laws of the $x^\mu$ and $\phi^a$ are independent of $\xi^m$. In other words, the transformation laws defined via the Maurer-Cartan form are naturally such that $x^\mu$ and $\phi^a$ form a representation of $G$. Let us put this in the context of the examples studied in the present paper. For the metafluid, the role of $\phi^a$ is played by $\sigma^i$ and, indeed, when looking at the transformation laws in \hyperref[app:TransformationLawsMF]{Appendix D 1}, the transformations of $\sigma^i$ and $x^\mu$ do not depend on any of the other fields. In other words, the metafluid is consistent with \cite{Klein:2017npd}. For the helical superfluid, $\left(\varpi^S\right)_\mu^{\hphantom{\mu}i}$ defined in \autoref{eq:HSFMCcoefs} contains a derivative of $\pi$, and so the constraints we impose do not satisfy the conditions stated above. However, from looking at  \hyperref[app:TransformationLawsHSF]{Appendix D 2}, if $\phi^a$ is played by $\pi$, then the transformation of $x^\mu$ and $\pi$ do not transform with respect to any of the other fields and so these also form a representation of $G$.

Then, in \cite{Klein:2017npd}, the authors claim that if one constructs an effective field theory for the massless NG modes, it will be equivalent to any method for eliminating the massive NG candidates since they drop out of the transformation laws for the massless NG modes. Our understanding of the argument is that if one assumes that the most general Lagrangian obtained with the Maurer-Cartan coefficients is equivalent to the most general Lagrangian built directly in terms of the fields, then, since the transformation laws are unaltered by the way we get rid of the massive NG modes, the final Lagrangians obtained via inverse Higgs constraints or by directly integrating out the mass should be equivalent. This assumes that the system formed by the inverse Higgs constraints is not overconstrained. To our knowledge, the assumption that the most general Lagrangian is the one built out of the Maurer-Cartan coefficients has not been proven. In fact, as we have seen for the metafluid, while $x^\mu$ and $\sigma^i$ form a representation of $G$, the theories obtained from imposing the inverse Higgs constraints and integrating out the masses are not equivalent. This is due to the fact that the inverse Higgs constraints overconstrain the system. On the other hand, for the helical superfluid, $x^\mu$ and $\pi$ not only form a representation of $G$, but also do not overconstrain the system. Yet, different theories were obtained from imposing three inverse Higgs constraints versus integrating out all of the masses. This suggests that the most general Lagrangian built out of the Maurer-Cartan form is not systematically equivalent to the most general Lagrangian built up directly in terms of the fields, and so the structure of the final Lagrangian can be sensitive to how we eliminate the massive NG candidates. Moreover, as noted before, when the homogeneous breaking of translations is involved, we could be led to situations where the system is overly constrained. It is not clear how to handle situations in which the inverse Higgs constraints relate the same $\phi^a$ to multiple $\xi^m$, leading to constraints solely on $\phi^a$. Although \cite{Klein:2017npd} does not \textit{a priori} consider the breaking of translations, we believe that the examples considered in this paper demonstrate that subtleties may be present in the extrapolation that the independence of the transformation laws for $\phi^a$ and $x^\mu$ from $\xi^m$ means the theory constructed directly from the transformation laws for $\phi^a$ is equivalent to any method chosen for eliminating the massive NG candidates.

Recall that we might lose generality in our computation when we chose a coset parametrization, when we chose a representation of the symmetry group, and when we use the Maurer-Cartan form to construct the building blocks of our EFTs. While choosing a different parametrization might lead to a different EFT, as long as we keep a parametrization of the form in \autoref{eq:Chosenparametrization}, we expect that the discussion about the pathological Inverse Higgs Constraints, both in the metafluid and helical superfluid, would not be affected. Indeed, choosing such a parametrization would always lead to massive terms of the form \eqref{eq:ExplicitMassiveTerms}. Hence, via the structure constants, the Inverse Higgs Constraints are always related to the algebra in the same way. For the helical superfluid, we have seen from the algebra that imposing the four Inverse Higgs Constraints would systematically lead to one field being expressed as a derivative of itself, which in turn lead to the pathological behavior. Concerning the metafluid case, this is not as systematic. Indeed, the algebra was such that we were obliged to impose all the Inverse Higgs Constraints. With more constraints to impose, there is a higher chances to get an overconstrained system and so, a pathology. As a future project, it would be nice to computationally check these assertions.

The second main result of this paper is that a large subset of theories found from these symmetry-breaking patterns have emergent enhanced shift symmetries, which lead to fractonic dispersion relations. For the metafluid, the final Lagrangian for the two massless fields $\sigma^i$, \autoref{eq:sigmaiEFT}, has an emergent enhanced shift symmetry when two of the coefficients in the final effective field theory are constrained to be equal, $\tilde{g} = \kappa_2$. As a result, one of the dispersion relations becomes a trivial mode $\omega_1^2 = 0$. To further demonstrate that there is a large class of theories that have this enhanced shift symmetry, we found that if we treat all the fields in the metafluid, $\sigma^i$, $\pi$, and $\chi$ as of the same order, as opposed to treating $\pi$ and $\chi$ as first order in derivatives of $\sigma^i$ as we did in \autoref{eq:MFLagrangian}, we would have obtained the Lagrangian in \autoref{eq:FullerMFLag}. There, we see that we can set two out of the 13 free parameters, $\kappa_4$ and $g_6$, to zero in order to obtain a Lagrangian that is only a function of $\pi$, $\chi$, $\partial_i \sigma^i$ and $\partial_t \sigma^i$. This has a symmetry under $\sigma^i \rightarrow \sigma^i + \epsilon^{ij}\partial_j f\left(x, y\right)$ for an arbitrary function $f$.

This raises an interesting possible outlook: while in these examples, only a few parameters need to be fixed, it would be worthwhile to explore the additional conditions to systematically connect the breaking of translations to fractonic behavior without fine tuning. We have good reason to believe that the breaking of translations is helpful for the generation of fractonic behavior as this has been done for symmetry-breaking patterns that do not involve dilations or shifts, e.g. \cite{Hirono:2021lmd} and \cite[eq. (21) upon setting $F = 0$]{Musso:2018}. That being said, it appears that the breaking of dilations and shifts in addition to translations was helpful in generating fractonic modes. 

We were also able to observe fractonic modes in the helical superfluid, both in the case where we imposed three inverse Higgs constraints and integrated out one massive mode and where we imposed no inverse Higgs constraints and integrated out four massive modes. It is interesting to note, however, that when we do utilize inverse Higgs constraints, we obtain the Lagrangian in \autoref{hsfEFTIHC}, where we need only set \textit{one} parameter to $0$ to obtain fractonic modes. On the other hand, without imposing inverse Higgs constraints, we obtain the theory, \autoref{eq:HSFFinalLag}, where more parameters would need to be fine tuned to obtain fractonic modes. It would be interesting to explore whether there is a connection between inverse Higgs constraints and the generation of fractonic modes. On a practical level, it is less computationally intensive to use inverse Higgs constraint than not and imposing more constraints on the dynamics could potentially impose constraints on mobility, a key feature of fractonic behavior. It would be useful to further explore whether inverse Higgs constraints can be used to generate fractonic theories more efficiently.

As an additional point of discussion, we would like to highlight that this work serves to give complimentary analysis to a question raised in \cite{Argurio:2021opr}, namely to what symmetry are the NG modes they find associated. They find that the actual physical modes, after diagonalizing the Lagrangian, are combinations of the perturbations around the background ground state. In the metafluid section, we find, in agreement with \cite{Argurio:2021opr}, that $\sigma^i$ are the NG modes associated to the breaking of shifts and that, in the low-energy limit far below the masses of the other NG candidates, the theory in \cite{Argurio:2021opr} is a theory contained in the generic metafluid theory found from the coset construction. Meanwhile, in the helical superfluid, we end up obtaining only one symmetry-protected massless mode, associated to the breaking of U(1), whereas in \cite{Argurio:2021opr}, there are two massless normal modes characterized by some mixing of the perturbations. The extra mode can arise from a theory in which one of the masses in \autoref{eq:HSFMassiveEFT} is set to zero from the start. This work indicates that it is possible one of the two modes found in \cite{Argurio:2021opr} is not actually a symmetry-protected massless NG mode.

The third main result of this paper is that we were able to develop a detailed and commented code for the community, found on GitHub \cite{SuppMat}, capable of handling coset constructions with complicated symmetry-breaking patterns. Indeed, the code in our case is able to produce and handle Lagrangians with well over 1000 terms. We find that even without rotations, it is still useful to use tensor packages, so long as you define appropriate projection vectors.

While we have already discussed a number of future directions related to the inverse Higgs constraint and fractonic physics, there are a number of future directions not already mentioned that we believe the models presented in this paper may be of some interest or use. First, unlike traditional inflationary models, \cite{Endlich_2013} developed a solid inflation model in which translations are broken but a diagonal combination of translations and shifts are unbroken. In future studies of solid inflation, it may be important to consider the subtleties of the coset construction studied in this paper. To do this, one would possibly need to embed this discussion in de Sitter space. Second, in \cite{Akyuz2024}, the authors extended the coset construction to the Schwinger-Keldysh formalism which allows for computations in open or finite-temperature systems. While the extension to Schwinger-Keldysh does not yet explicitly include the breaking of spacetime translations, it may be worthwhile to use it to explore our computations in a way that more closely resembles a laboratory setup. Alternatively, one could study the sensitivity of physical observables to the inequivalence of the obtained Lagrangians. To do this, one would need to go to higher order in fields to study interactions and therefore amplitudes. Third, one could use the bottom-approach of holography to describe the homogeneous breaking of translations \cite{Donos:2013eha, Ling:2014laa}. This will allow us to explore more exotic QFTs, and possibly probe whether we still obtain fractonic behavior. Finally, we could extend this approach to the inhomogeneous breaking of translations and see if fractonic behaviors emerge as well. As an example, \cite{Brauner:2024juy} studied the kink instanton and showed that the NG mode associated to the spontaneous symmetry-breaking of translations at the level of the domain wall can only propagate in the transverse direction of the domain wall. Intuitively, this reduced mobility is due to the fact the momentum is no longer defined in the longitudinal direction. It would be interesting to further explore to what extent the breaking of translations lead to fractonic behavior.

\begin{acknowledgments}
We wish to thank Alberto Nicolis and Enrico Pajer for many useful discussions and guidance throughout the project as well as Carlos Hoyos, Daniele Musso, and Riccardo Argurio for their feedback on the draft of this paper. A.C. is supported by the National Science Foundation Graduate Research Fellowship under Grant No. DGE-2036197. D.N. expresses his gratitude to the Belgian American Educational Foundation, Wallonie-Bruxelles International, and Vocatio for funding his research residency at Columbia University in the City of New York during the early phase of this project. He is also grateful to the Wiener-Anspach Foundation for supporting his current postdoctoral stay at the University of Cambridge. J.S. acknowledges support from the DOE Grant No. DE-SC011941.
\end{acknowledgments}

\newpage

\appendix

\section{\label{app:CosetConstruction} Review of the coset construction for spacetime symmetries}

In this appendix, we provide a concise review of the coset construction, a method for constructing an action for NG candidates solely on the basis of symmetry considerations \cite{Coleman:1969sm,Callan:1969sn,Salam:1969rq,Volkov:1973vd,Ogievetsky:1974,DHoker:1994rdl}. NG candidates arise due to the spontaneous breaking of a symmetry group $G$ to a subgroup $H$, by which we mean that a particular ground state may not be invariant under all the symmetry transformations of $G$, but only a particular subset, $H$. The action for these NG candidates must still be invariant under the full symmetry group, but the latter will be realized linearly for the unbroken symmetries and nonlinearly for the broken symmetries. As an example, consider a complex field, $\phi\left(x\right)$ whose ground state breaks U(1). If we write
\begin{equation}
\label{eq:phirho}
\phi\left(x\right) = \rho\left(x\right) e^{i\pi\left(x\right)},
\end{equation}
then if we apply a U(1) transformation, the NG candidate $\pi\left(x\right)$ will inherit a shift transformation, which is nonlinear. Since symmetry-breaking occurs below a certain energy scale, the coset construction is particularly useful in the writing of EFTs. 

The coset construction provides invariant building blocks by defining a local parametrization of the coset space, $G/H_0$, where $H_0$ is the subgroup $H$ without the unbroken translations \footnote{Note the implicit assumption that $H_0$ itself is a continuous (or trivial) subgroup of $G$.}, introducing NG candidate fields, and establishing a nonlinear realization of the full symmetry group $G$, where unbroken symmetries are realized linearly, and broken symmetries are realized nonlinearly. Before delving into the details, we begin with some essential definitions: $\bar{P}_\mu$ is defined as the unbroken spatial translation generator, $T_A$ denotes the unbroken generators, and $X_a$ are the generators of the broken symmetries. As an example, say $G$ contains translations generated by $P_\mu$, U(1) generated by $Q$, and rotations generated by $J$. If, after symmetry-breaking $P_0 - k Q$ and $J$ are unbroken, we would say $H_0$ contains only rotations, $\bar{P}_0 = P_0 - kQ$ is an unbroken translation, and $G/H$ is just U(1).

Inspired by \autoref{eq:phirho}, we define the coset parametrization as
\begin{equation}
\label{eq:Sumparametrization}
U(x,\pi(x)) = e^{ix^\mu \bar{P}_\mu} e^{i\pi^a(x) X_a},
\end{equation}
with $\pi^a\left(x\right)$ being the NG candidates, which parametrize fluctuations in the direction of the broken generators. The operator $e^{ix^\mu \bar{P}_\mu}$ in the parametrization has been introduced so that the fields have standard transformations under translations. This parametrization is not unique. In fact, the parametrization we use in the main text of the paper includes products of exponentials,
\begin{equation}
\label{eq:Chosenparametrization}
U\left(x, \pi\left(x\right)\right) = e^{ix^\mu \bar{P}_\mu} e^{i\pi^1\left(x\right)X_1} e^{i\pi^2\left(x\right)X_2} \dots.
\end{equation} 
For the purposes of this section, we stick with the parametrization in \autoref{eq:Sumparametrization}, which will not change the conclusions of this appendix. This flexibility implies that while the coset construction for spacetime symmetries yields a very general EFT, it does not necessarily provide the most general theory. Specifically, some degree of generality is sacrificed at the level of the chosen parametrization because not all parametrizations are equivalent \cite{McArthur:2010zm,Creminelli:2014zxa,Finelli:2019keo}. Moreover, to the best of our knowledge, there is no proof that the most general Lagrangian constructed from covariant building blocks, which, as we will see, will be functions of the NG candidates, corresponds to the most general Lagrangian expressible directly in terms of the NG candidate fields.

The fundamental assumption of the coset construction is that the vector space generated by the broken generators and the vector space generated by the unbroken translation generators respectively form a representation of the subgroup $H_0$. At the level of the Lie algebra, this can be expressed in terms of the unbroken translation generators, the broken generators, and the unbroken generators with the following form:
\begin{equation}
\label{eq:FundAssump}
\comm{X_a}{T_A} = i f_{aA}^{\hphantom{aA}b} X_b \qquad \text{and} \qquad \comm{\bar{P}}{T_A} = if_{\mu A}^{\hphantom{\mu A}\nu} \bar{P}_\nu.
\end{equation}
As an example, if $T_A$ is the rotation generator, $J_i$, then $X_j$ is a vector representation of rotations if the commutator of $X_j$ and $J_i$ is $i\epsilon_{ij}^{\hphantom{ij}k}X_k$. 

To define suitable transformation laws, which will be nonlinear for broken symmetries and linear for unbroken ones, we need to inspect the action of a group element on the coset parametrization. Since the element of the coset space is an element of the group, multiplication by a group element, $g$, will yield another element of the coset space up to an element of $H_0$, $h$,
\begin{equation}
\label{eq:Pushingh}
g U(x, \pi(x)) = U(\tilde{x},\tilde{\pi}(\tilde{x})) h\left(\pi, g, x\right),
\end{equation}
where $\tilde{x}$ and $\tilde{\pi}^a$ are the transformed coordinates and fields respectively. In practice, one would write the elements of the group, $g$ and $h$, as exponentials of the generators and use the Baker-Campbell-Hausdorff formulas in \autoref{eq:BCH1}--\autoref{eq:BCH3} and the commutators in \autoref{eq:FundAssump} to write the transformation laws in exactly this form. Note that while $g$ and $h$ are generic group elements, $U$ is expressed in terms of Lie algebra elements. This does not lead to any loss of generality, since it is defined for small fields and so $U$ is, by definition, continuously connected to the identity. For a comprehensive discussion, we refer the reader to \cite{NaegelsThesis}. 

An object that is useful for building invariants is the Maurer-Cartan 1-form, $U^{-1} \partial_\mu U$, which can in general be written as a linear combination of the generators in $G$, as it is a Lie algebra valued element: 
\begin{equation}
\label{eq:GenericMCForm}
U^{-1}\partial_\mu U = i\left(e_\mu^{\hphantom{\mu}\nu} \bar{P}_\nu + e_\mu^{\hphantom{\mu}\nu} \varpi_\nu^{\hphantom{\nu}a} X_a - \mathcal{A}_\mu^{\hphantom{\mu}A}T_A\right),
\end{equation}
where $e$, $\varpi$, and $\mathcal{A}$ are all functions of the spacetime coordinates, $x$, and the NG candidates, $\pi$.

By applying the transformation rule and using the fundamental assumption in \autoref{eq:FundAssump}, the transformation laws for the Maurer-Cartan coefficients under the group $G$ are as follows \cite{NaegelsThesis}:
\begin{align}
e_\mu^{\hphantom{\mu}\alpha}(\tilde{x}, \tilde{\pi}) &= \frac{\partial x^\nu}{\partial \tilde{x}^\mu} e_\nu^{\hphantom{\nu}\beta}(x, \pi) h_\beta^{\hphantom{\beta}\alpha}(g, x, \pi), \\
\varpi_\alpha^{\hphantom{\alpha}a}(\tilde{x}, \tilde{\pi}) &= \left( h^{-1}(g, x, \pi) \right)_{\alpha}^{\hphantom{\alpha}\nu} \varpi_\nu^{\hphantom{\nu}b}(x, \pi) h_b^{\hphantom{b}a}(g, x, \pi), \\
\mathcal{A}_\mu^{\hphantom{\mu}A}(\tilde{x}, \tilde{\pi}) T_A &= \mathcal{A}_\mu^{\hphantom{\mu}A}(x, \pi) h\left(g, \pi, x\right) T_A h\left(g, \pi, x\right)^{-1} + i h\left(g, \pi, x\right) \partial_\mu h^{-1}\left(g, \pi, x\right),
\end{align}
where $h_\mu^{\hphantom{\mu}\nu}\left(g, \pi, x\right)$ and $h_b^{\hphantom{b}a}\left(g, x, \pi\right)$ are representations of the function $h\left(g, \pi, x\right) \in H_0$, introduced in \autoref{eq:Pushingh}, much in the way the rotation matrix is a representation of an SO(3) element. Note that $h_\mu^{\hphantom{\mu}\nu}\left(g, \pi, x\right)$ and $h_b^{\hphantom{b}a}\left(g, x, \pi\right)$ is a function of a generic group element, $g$, and so the full symmetry group $G$ is represented by a linear representation of $H_0$. First, the transformation law of $e_\mu^{\hphantom{\mu}\alpha}$ is that of a tetrad. In other words, if we wanted to define an invariant integration measure, we should use $\det\left(e\right)\dd[3]{x}$ instead of $\dd[3]{x}$. Next, the Maurer-Cartan coefficients transform covariantly under the full symmetry group G. As a result, we can built invariant objects by just contracting indices of $H_0$. Finally, $\mathcal{A}_\mu^{\hphantom{\mu}A}$ transforms as a gauge field under the unbroken subgroup. This leads to a natural definition of the covariant derivative
\begin{equation}
\label{eq:GeneralCovD}
\nabla_\mu = \left(e^{-1}\right)_\mu^{\hphantom{\mu}\nu}\left(\partial_\nu - i \mathcal{A}_\nu^{\hphantom{\nu}A}T_A\right).
\end{equation}
\vspace{-1.5cm}

Therefore, if we are to construct an invariant action, we can use the Maurer-Cartan coefficients and their covariant derivatives and contract the indices in such a way that makes invariance under the unbroken subgroup manifest and integrate over the invariant integration measure. In other words,
\begin{equation}
\label{eq:GenericAction}
S = \int \mathcal{L}\left(\varpi_\mu^{\hphantom{\mu}a}(x, \pi), \nabla_\alpha\varpi_\mu^{\hphantom{\mu}a}(x, \pi), \dots \right) \det\left(e\left(x, \pi\right)\right) \dd[3]{x}.
\end{equation}
This action will be invariant under the full symmetry group $G$. In general, this action could include terms that transform up to a total derivative, which are known as Wess-Zumino terms. While they are not relevant for our paper, they are further discussed in \cite{DHoker:1994rdl}. Since there is spontaneous symmetry-breaking, there is a vacuum expectation value which can be considered to be a cutoff. So long as the momentum is smaller than this cutoff, we can expand this action in terms of powers of the momentum over the cutoff. When we are dealing with only massless modes, this will be a derivative expansion. For a more comprehensive and technical discussion of the coset construction, refer to \cite{NaegelsThesis} and \cite{Penco:2020kvy}.

\section{Review of the inverse Higgs constraint}
\label{app:IHCReview}

When spacetime symmetries are broken, not every NG candidate that appears in the action, \autoref{eq:GenericAction}, will necessarily be massless. In this appendix, we will present an operational prescription which will eliminate (some of) these massive NG candidates. This is accomplished by expressing them in terms of massless NG candidates in a symmetrically consistent way. These relationships are called inverse Higgs constraints. A more formal definition will be provided once the technicalities have been reviewed.

If we expand the building blocks to linear order in the fields, we find \cite[eqs. 5.2.28 - 5.2.30]{NaegelsThesis},
\begin{align}
e_\mu^{\hphantom{\mu}\nu} &\approx \delta_\mu^{\hphantom{\mu}\nu} - \pi^a f_{\mu a}^{\hphantom{\mu a}\nu} + \mathcal{O}\left(\pi^2\right),\\
\label{eq:MCExpand} \varpi_\mu^{\hphantom{\mu}a} &\approx \partial_\mu \pi^a - \pi^b f_{\mu b}^{\hphantom{\mu b}a} + \mathcal{O}\left(\pi^2\right),\\
A_\mu^{\hphantom{\mu}A} &\approx \pi^a f_{\mu a}^{\hphantom{\mu a} A} + \mathcal{O}\left(\pi^2\right),
\end{align}
where $f$ with any indices are the structure constants of the Lie algebra. There are terms that do not contain any derivatives and so nonderivative coupling terms, such as massive terms, can appear in the Lagrangian. This is due to the fact that certain structure constants, $f_{\mu a}^{\hphantom{\mu a}\dots}$, are not $0$ \cite{Rothstein:2017twg}, which implies that the commutators $\comm{\bar{P}_\mu}{X_a}$ are not zero. One of the issues that arises as a result of these terms is that we can no longer expand in powers of derivatives, as terms like $\pi^2$ will have a dimensionful coupling constant that may not have the same scale as a derivative. The fields, $\pi^a$, that appear with no derivatives and have mass term, $\left(\pi^a\right)^2$, are either massive fields or spurious, meaning their equations of motion are purely algebraic. To understand the origin of these non-NG modes requires some knowledge about the UV theory. However, there are arguments that interpret the spurious fields as a type of gauge redundancy \cite{Low:2001bw, Watanabe:2013iia, McArthur:2010zm, Nicolis:2013sga, Endlich:2013vfa}. Additionally, inspecting the transformation laws yields relationships between the Noether currents associated to different broken symmetries. In turn, at least when translations are unbroken, this establishes a relationship between the symmetry group, $G$, and the spectrum of NG modes \cite{Brauner:2014aha}. Note that when only internal symmetries are broken, $f_{\mu a}^{\hphantom{\mu a}\dots} = 0$ and so there would no longer be any explicit mass terms.

For our purposes, we will only focus on theories that are up to quadratic order in the fields. As such, the no-derivative terms that we are interested in are massive terms. Additional terms linear in the field can always be reabsorbed by completing the square and performing a field redefinition. In order for there to be a mass term $\left(\pi^a\right)^2$, we could have
\begin{equation}
\label{eq:ExplicitMassiveTerms}
\begin{aligned}
\varpi_\mu^{\hphantom{\mu}a}\varpi_\nu^{\hphantom{\nu}b} &\supset f_{\mu c}^{\hphantom{\mu c} a} f_{\nu d}^{\hphantom{\nu d}b}\pi^c \pi^d,\\
\nabla_\mu \varpi_\nu^{\hphantom{\nu}a} &\supset -f_{\nu b}^{\hphantom{\nu b}a}f_{\mu c}^{\hphantom{\mu c}A}  \pi^b \pi^c T_A.
\end{aligned}
\end{equation}
Note that we are not yet contracting any indices as we have not specified $H_0$. Since $f_{\mu c}^{\hphantom{\mu c}A}$ is multiplying $f_{\nu b}^{\hphantom{\nu b}a}$, then if $f_{\nu b}^{\hphantom{\nu b}a}$ is $0$ that massive term will not appear. Moreover, when $f_{\nu b}^{\hphantom{\nu b}a} f_{\mu c}^{\hphantom{\mu c}A}$ leads to a mass term, that mass term is already counted in $\varpi_\nu^{\hphantom{\nu}a}\varpi_\mu^{\hphantom{\mu}d}$. There are other mass terms that can arise, including from the second-order expansion of $\det\left(e\right)$ and combinations of $\det\left(e\right)$ with terms linear in the Maurer-Cartan coefficient that will be linked to $f_{\mu a}^{\hphantom{\mu a}\nu}$. Therefore, in order to have massive terms, we require, at minimum, that $f_{\mu a}^{\hphantom{\mu a}b} \neq 0$ and/or $f_{\mu a}^{\hphantom{\mu a}\nu} \neq 0$. At the level of the algebra, this means
\begin{equation}
\label{eq:IHCCommutator}
\comm{\bar{P}_\mu}{X_a} \supset X_b \qquad \text{and/or} \qquad \comm{\bar{P}_\mu}{X_a} \supset \bar{P}_\nu.
\end{equation}
These massive modes will be associated to the broken generator $X_a$. This suggests that by looking at the algebra, we can already predict an \textit{a priori} counting of the number of massive modes\footnote{From \autoref{eq:ExplicitMassiveTerms} and the comment on $\det\left(e\right)$, the structure constants of the algebra can as well provide an indication on the value of the masses. See for example \cite{Nicolis:2013sga}, where they study how the NG candidates' gaps scale with the chemical potential when the latter is introduced via an homogeneous breaking of time translation.}. Specifically, the counting of terms that will have a mass term $\left(\pi^a\right)^2$ is equal to the number of distinct generators $X_a$ that appear in commutators like \autoref{eq:IHCCommutator}. For example, by looking at the algebra in \autoref{eq:MFLieAlgebra}, there are four commutators of this form and so there are four mass terms in the Lagrangian in \autoref{eq:HSFMassiveEFT}-- $\left(\sigma^1\right)^2$, $\left(\sigma^2\right)^2$, $\theta^2$, and $\chi^2$. That there are massive fields is a nondynamical reduction of NG modes compared to NG candidates. Note that there may still yet be a further dynamical reduction of NG modes compared to NG candidates so this does not provide a true counting of the number of NG modes, as discussed around \autoref{eq:ConjugationReduction}.\footnote{Another reduction could occur if there are terms both linear and quadratic in time derivatives. Finding the dispersion relations would result in a massive mode due to the competition between these two terms.}

We would like to derive an effective field theory for only the massless NG modes using the coset construction. However, by the above arguments, there will be massive fields depending on the algebra. One way to handle the massive modes is to integrate out the massive modes. An alternative would be to write an effective field theory that is fully invariant under the symmetry group with only the NG candidates that are \textit{a priori} massless according to the algebra. To do this, we would need to express the NG candidates that are massive in terms of \textit{a priori} massless NG candidates. One naive way to do this is by setting one of the Maurer-Cartan coefficients leading to the massive mode to $0$,
\begin{equation}
\label{eq:SoltoIHC}
\varpi_\mu^{\hphantom{\mu}a} = 0 \Rightarrow \partial_\mu \pi^a \approx \pi^b f_{\mu b}^{\hphantom{\mu b}a}.
\end{equation}
However, this will potentially spoil the symmetry as $\varpi_\mu^{\hphantom{\mu}a}$ transforms covariantly. For example, if $\pi^a$ and $\pi^b$ transform together as a vector under $H_0 =$ SO(2), we cannot set a constraint on only one of the components of that vector. Another way would be to set every Maurer-Cartan coefficient to $0$, but this will trivialize the theory. Instead, we would like to find a way to impose some components to zero, without spoiling the symmetry and ensuring that these imposed constraints are able to be solved algebraically, as in \autoref{eq:SoltoIHC}. 

In order to accomplish that goal, we need two assumptions.  First, the set of $X_a$ and the set of $\bar{P}_\mu$ respectively form a completely reducible representation of $H_0$. This can be checked at the level of the algebra. This, in turn, means that $\varpi_\mu^{\hphantom{\mu}a}$ is a completely reducible representation of $H_0$. This means we can impose a subset of the components of $\varpi_\mu^{\hphantom{\mu}a}$ that form an irreducible representation of $H_0$ to $0$. This ensures that we do not trivialize the theory while also remaining consistent with the symmetries of the theory. Going back to the previous example where $H_0 =$ SO(2), so we have $\varpi_\mu^{\hphantom{\mu}i}$ which can be reduced to the spatial vector and tensor $\varpi_0^{\hphantom{0}i}$ and $\varpi_j^{\hphantom{j}i}$. These two are separate irreducible representations of SO(2), a vector and a tensor. Second, we would need to solve the constraint algebraically. Two necessary, but not sufficient, conditions to be able to solve these constraints algebraically are that each component $\varpi_\mu^{\hphantom{\mu}a}$ contains a linear term in $\pi^b$, i.e. $f_{\mu b}^{\hphantom{\mu b}a} \neq 0$ as in \autoref{eq:MCExpand}, and that it does not depend on the derivative of $\pi^b$.\footnote{This condition implies extra conditions on the structure constants. This does not affect our results since we remain at quadratic order in the fields. For more details about these conditions see \cite{Klein:2017npd}} These conditions are not sufficient as the different obtained constraints, despite being individually algebraic in terms of $\pi^b$, may mix, leading to possible nonalgebraic relationships among NG candidates due to their respective $\partial_\mu \pi^a$ dependency. An example of such situation is discussed in the next paragraph. The act of setting one component of $\varpi_\mu^{\hphantom{\mu}a}$ to $0$ is called an inverse Higgs constraint \cite{Ivanov:1975zq}. This should be done in a way that is consistent with the aforementioned assumptions, which may force one to impose several inverse Higgs constraints. The goal of imposing these constraints is to eliminate the explicit massive modes of the effective field theory after spontaneous symmetry-breaking.

There are multiple subtleties that can occur when trying to impose inverse Higgs constraints. This is worth clarifying since the usual procedure that does work most of the time is to identify the possible constraints from the algebra and impose them immediately. One subtlety occurs if for one $X_a$ there are multiple commutators that could give inverse Higgs constraints, say $\comm{P_\mu}{X_a} \supset X_b$ and $\comm{P_\mu}{X_a} \supset X_c$. Such a situation arises in \cite{Rothstein:2017twg} and is most easily handled by simply choosing one of the two commutators to source the constraint. However, we illustrate in this paper how this may give rise to additional subtleties. For example, imposing only one of the two constraints might violate the underlying symmetry $H_0$ or, even when that does not happen, one might still run into a situation where the inverse Higgs constraint does not yield a low-energy theory that is equivalent to integrating out the masses. Other subtleties include choices in how we parametrize our coset space. For example, we may choose to write $U$ as a product of exponentials, $e^{ix^\mu\bar{P}_\mu}e^{i \pi^1 X_1} e^{i\pi^2 X_2}\dots$ or have the sum in the argument of the exponential, $e^{ix^\mu\bar{P}_\mu} e^{i\pi^1 X_1 + i \pi^2 X_2}$. This may affect the ease with which we impose inverse Higgs constraints, as further discussed \cite{Klein:2017npd}.

The current paradigm in the literature is that, so long as we do not break spacetime translations and there is a one-to-one matching between the possible inverse Higgs constraints and the fields which can be eliminated by these same constraints, there is no loss of generality by imposing the inverse Higgs constraints to solely study the massless fields \cite{McArthur:2010zm, Brauner:2014aha, Klein:2017npd}. Said another way, if we have a massive field, imposing the associated inverse Higgs constraint would be equivalent to integrating that mass out. If the equation of motion for a field is algebraic, then using the associated inverse Higgs constraint would be equivalent to using that equation of motion. To be clear, the effective field theories obtained are called equivalent if there is a field redefinition that maps one to the other \cite{Finelli:2019keo}. 

\section{\label{app:ImportantIdentities} Useful identities}
There are a few important corollaries to the Baker-Campbell-Hausdorff formula that are essential for computing transformation laws and Maurer-Cartan coefficients. The first tells you how an operator $Y$ transforms under a similarity transformation associated with $X$ \citep[Prop. 3.35]{Hall2015}:
\begin{equation}
\label{eq:BCH1}
    e^X Y e^{-X} = Y + \comm{X}{Y} + \frac{1}{2!}\comm{X}{\comm{X}{Y}} + \dots = \sum_{n = 0}^{\infty} \frac{\comm{\left(X\right)^n}{Y}}{n!},
\end{equation}
where $\comm{\left(X\right)^0}{Y} \equiv Y$ and, for $n \neq 0$,
\begin{equation}
\comm{\left(X\right)^n}{Y} \equiv [\underbrace{X, \cdots [X, [X}_{\text{$n$ times}}, Y]] \cdots ].
\end{equation}
The next corollary will allow us to pass an exponential $e^X$ through another one, $e^Y$ \citep[pg. 25]{Rossmann2002}:
\begin{equation}
\label{eq:BCH2}
    e^X e^Y = \exp\left(Y + \comm{X}{Y} + \frac{1}{2!}\comm{X}{\comm{X}{Y}} + \dots\right) e^X = \exp\left(\sum_{n = 0}^{\infty} \frac{\comm{\left(X\right)^n}{Y}}{n!}\right) e^X.
\end{equation}
Lastly, we may have to untangle an exponential with a sum and make it a product of exponentials to recover the correct coset parametrization. This is done using the Zassenhaus formula \citep[Section IV.]{Magnus1954}: 
\begin{equation}
\label{eq:BCH3}
    e^{X + Y} = e^X e^Y \exp\left(-\frac{1}{2!}\comm{X}{Y}\right)\exp\left(\frac{1}{3!}\left(2\comm{Y}{\comm{X}{Y}} + \comm{X}{\comm{X}{Y}}\right)\right) \dots.
\end{equation}
This identity does not have a simple closed form as the other two did.

\section{\label{app:TransformationLaws} Transformation laws}
As noted in the main text, one way to construct the Lagrangian after specifying the parametrization of the coset space is to look at how the NG candidates transform and write down every possible operator that respects those transformation laws. The latter is not so simple in practice, which is why the approach of using the Maurer-Cartan form is so often adopted. However, knowing the transformation laws has two benefits. One, it is a good consistency check as whatever action we write down must be invariant under these transformations. Two, it gives yet another perspective from which to study the inverse Higgs constraint, as further discussed in Sec. \autoref{s:Discussion}. We can compute the transformation laws for the NG candidates, $\pi^a$, under a generic group element $g$ by studying its effect on the coset parametrization $U\left(x, \pi^a\right)$ \citep[eq. 5.2.9]{NaegelsThesis},
\begin{equation}
g U\left(x, \pi^a\right) = U\left(\tilde{x}, \tilde{\pi}^a\right) e^{i u^A\left(g, \pi\left(x\right)\right)T_A},
\end{equation}
where $T_A$ are the unbroken generators, whose transformations are parametrized by $u^A$.

\subsection{Metafluid}
\label{app:TransformationLawsMF}
Let us recall that the coset space for the metafluid was parametrized from \autoref{eq:EquivMFparametrization}
\begin{equation}
U\left(x, \psi^i, \pi, \chi\right) = e^{i x^\mu P_\mu} e^{i\psi^i S_i} e^{i\pi Q} e^{i\chi D},
\end{equation}
where $\psi^i \equiv \sigma^i + kx^i$. Crucially, the full symmetry group not only includes shifts, U(1) transformations, and dilations, but also unbroken translations and rotations. In the following computations, we will use the Lie algebra in \autoref{eq:FullLieAlg}.

Beginning with shifts parametrized by $s^i$, we find that since shifts commute with momentum, we get a shift of $\psi^i$,
\begin{equation}
e^{i s^i S_i} U\left(x, \psi^i, \pi, \chi\right) = e^{i x^\mu P_\mu} e^{i\left(\psi^i + s^i\right) S_i} e^{i\pi Q} e^{i\chi D}.
\end{equation}
Therefore, under shifts, everything remains the same except the shift NG candidate, whose transformation is just a shift
\begin{equation}
\label{eq:MFShifttrans}
\sigma'^{\:i} = \sigma^i + s^i. 
\end{equation}

Let us now apply a U(1) transformation, parametrized by $\alpha$. From the Lie algebra in \autoref{eq:FullLieAlg}, $Q$ commutes with everything except $S_i$. Therefore, we need only study the following combination,
\begin{equation}
\label{eq:RotationAnalogy}
\begin{aligned}
e^{i\alpha Q} e^{i \psi^i S_i} &= \exp\left(i\psi^i S_i + \comm{i\alpha Q}{i \psi^i S_i} + \frac{1}{2} \comm{i\alpha Q}{\comm{i\alpha Q}{i \psi^i S_i}} + \dots\right) e^{i\alpha Q},\\
&= \exp\left(i\psi^i S_i + i\alpha \psi^i \epsilon_i^{\hphantom{i}j}S_j - \frac{i\alpha^2}{2}\psi^i S_i + \dots\right)e^{i\alpha Q},\\
&= e^{i\psi^i R_i^{\hphantom{i}j}\left(\alpha\right)S_j} e^{i\alpha Q}. 
\end{aligned}
\end{equation}
On the right-hand side of the first line we used the Baker-Campbell-Hausdorff identity in \autoref{eq:BCH2}. On the second line, we evaluated the commutators using \autoref{eq:FullLieAlg} and, on the third line, we noted that the expansion was equivalent to that of a rotation matrix. As an aside, by noting the similarity of the identities \autoref{eq:BCH2} and \autoref{eq:BCH1}, we can also write
\begin{equation}
\label{eq:MChelp1}
e^{-i\alpha Q} S_i e^{i\alpha Q} = R_i^{\hphantom{i}j}\left(-\alpha\right) S_j.
\end{equation}
Therefore, under U(1) transformations, $U$ transforms as
\begin{equation}
e^{i\alpha Q} U\left(x, \psi^i, \pi, \chi\right) = e^{i x^\mu P_\mu} e^{i R^i_{\hphantom{i}j}\left(-\alpha\right) \psi^j S_i} e^{i\left(\pi + \alpha\right) Q} e^{i\chi D},
\end{equation}
where we used the fact that the transpose of a rotation matrix is its inverse. As a result, everything remains the same, except the U(1) NG candidate, whose transformation is a shift, and $\sigma^i$, whose transformation is a rotation and translation:
\begin{equation}
\label{eq:MFU1trans}
\sigma'^{\:i} = R^i_{\hphantom{i}j}\left(-\alpha\right) \left(\sigma^j + k x^j\right) - k x^i \qquad \text{and} \qquad \pi' = \pi + \alpha.
\end{equation}

Next, we will study the transformation of $U$ under dilations, parametrized by $\lambda$. First, using \autoref{eq:BCH2},
\begin{equation}
\begin{aligned}
e^{i \lambda D} e^{i x^\mu P_\mu} &= \exp\left(i x^\mu P_\mu + \comm{i\lambda D}{i x^\mu P_\mu} + \frac{1}{2} \comm{i\lambda D}{\comm{i\lambda D}{i x^\mu P_\mu}} + \dots\right) e^{i \lambda D},\\
&= \exp\left(i x^\mu P_\mu - i \lambda x^\mu P_\mu + \frac{i}{2}\lambda^2 x^\mu P_\mu + \dots\right) e^{i\lambda D},\\
&= e^{i e^{-\lambda} x^\mu P_\mu} e^{i\lambda D}.
\end{aligned}
\end{equation}
Next, noting that $\comm{D}{S_i} = -i\Delta S_i$, which differs by a factor of $-\Delta$ from $\comm{D}{P_\mu}$, 
\begin{equation}
e^{i\lambda D} e^{i\psi^i S_i} = e^{i e^{\Delta\lambda} \psi^i S_i} e^{i\lambda D}.
\end{equation}
Once again, as an aside, noting the similarity between \autoref{eq:BCH2} and \autoref{eq:BCH1}, we can say
\begin{equation}
\label{eq:MChelp2}
e^{-i\lambda D} P_\mu e^{i\lambda D} = e^{\lambda} P_\mu \qquad \text{and} \qquad e^{-i\lambda D} S_i e^{i\lambda D} = e^{-\Delta\lambda} S_i.
\end{equation}
Therefore, under dilations, we find
\begin{equation}
\label{eq:MFUnderD}
e^{i\lambda D} U\left(x, \psi^i, \pi, \chi\right) = e^{ie^{-\lambda} x^\mu P_\mu} e^{i e^{\Delta \lambda} \psi^i S_i} e^{i\pi Q} e^{i\left(\chi + \lambda\right)D}.
\end{equation}
In other words, the $\pi$ field remains unchanged and the remaining fields and coordinates transform as follows:
\begin{equation}
\label{eq:MFDiltrans}
x'^\mu = e^{-\lambda} x^\mu, \qquad \sigma'^{\:i} = e^{\Delta \lambda} \left(\sigma^i + k x^i\right) - k e^{-\lambda} x^i, \qquad \text{and} \qquad \chi' = \chi + \lambda.
\end{equation}

For completeness, we also note how the fields and coordinates transform under unbroken translations and rotations, $\bar{P}_\mu$ and $\bar{J}$. In these cases, it is easier to use the original coset parametrization,
\begin{equation}
U\left(x, \sigma^i, \pi, \chi\right) = e^{ix^\mu \bar{P}_\mu} e^{i\sigma^i S_i} e^{i\pi Q} e^{i\chi D}.
\end{equation}
Under unbroken translations parametrized by $a^\mu$, we simply find that the fields remain unchanged and
\begin{equation}
\label{eq:MFtrans}
x'^\mu = x^\mu + a^\mu.
\end{equation}
Unbroken rotations commute with everything except shifts and translations, where the commutator has the same exact structure as $\comm{Q}{S_i}$ in \autoref{eq:MFLieAlgebra}. We can therefore write down the transformation laws by analogy to \autoref{eq:RotationAnalogy}, telling us that $\pi$ and $\chi$ remain unchanged while
\begin{equation}
\label{eq:MFRottrans}
x'^{\:i} = R^i_{\hphantom{i}j}\left(-\varphi\right) x^j \qquad \text{and} \qquad \sigma'^{\:i} = R^i_{\hphantom{i}j}\left(-\varphi\right)\sigma^j,
\end{equation}
where the unbroken rotations are parametrized by $\varphi$. 

To summarize, the nontrivial transformation laws of the coordinates and fields under the full symmetry group are given by \autoref{eq:MFShifttrans}, \autoref{eq:MFU1trans}, \autoref{eq:MFDiltrans}, \autoref{eq:MFtrans}, and \autoref{eq:MFRottrans}.

\subsection{Helical superfluid}
\label{app:TransformationLawsHSF}
In the case of the helical superfluid, we used the parametrization from \autoref{eq:HSFparametrization},
\begin{equation}
U\left(x, \psi, \sigma^i, \chi, \theta\right) = e^{ix^\mu P_\mu} e^{i\psi Q} e^{i\sigma^i S_i} e^{i\chi D} e^{i\theta J},
\end{equation}
where $\psi \equiv \pi + kx$. In this case, all symmetries are broken by the ground state except unbroken translations. In this parametrization, we can once again make use of the algebra in \autoref{eq:FullLieAlg}.

Beginning with U(1) transformations parametrized by $\alpha$, we find
\begin{equation}
e^{i\alpha Q} U\left(x, \psi, \sigma^i, \chi, \theta\right) = e^{ix^\mu P_\mu} e^{i\left(\psi + \alpha\right)Q} e^{i\sigma^i S_i} e^{i\chi D} e^{i\theta J}.
\end{equation}
Therefore, under U(1) transformations, the only transformation is a shift
\begin{equation}
\label{eq:HSFU1trans}
\pi' = \pi + \alpha.
\end{equation}
We will continue to write the transformation laws on the true NG candidate, $\pi$, as opposed to $\psi$. 

Next, we will study shifts parametrized by $s^i$. Since shifts commute with themselves, the expansion in \autoref{eq:BCH2} truncates at first order, giving
\begin{equation}
\begin{aligned}
e^{is^i S_i} e^{i\psi Q} &= \exp\left(i\psi Q + \comm{i s^i S_i}{i \psi Q}\right) e^{i s^i S_i},\\
&= \exp\left(i\psi Q - i s^i \psi \epsilon_i^{\hphantom{i}j} S_j\right) e^{i s^i S_i}.
\end{aligned}
\end{equation}
Now we will need to untangle this exponential into a product of exponentials using \autoref{eq:BCH3}. Since shifts commute with each other, this formula reduces significantly
\begin{equation}
\begin{aligned}
e^{i s^i S_i} e^{i\psi Q} &= e^{i \psi Q} e^{is^i S_i} e^{- i s^i \psi \epsilon_i^{\hphantom{i}j} S_j} \\
&\hspace{0.5cm} \times \exp\left(-\frac{1}{2!}\comm{i\psi Q}{-i s^i \psi \epsilon_i^{\hphantom{i}j}S_j}\right)\exp\left(\frac{1}{3!}\comm{i\psi Q}{\comm{i\psi Q}{-i s^i \psi \epsilon_i^{\hphantom{i}j}S_j}}\right)\dots.
\end{aligned}
\end{equation}
Since the commutator of $Q$ with $S_i$ gives us $S_i$, we were free to move $e^{is^i S_i}$ to the front. Additionally, every term is part of a Taylor series expansion of a rotation matrix. Therefore, the above expression simplifies to
\begin{equation}
e^{is^i S_i} e^{i\psi Q} = e^{i\psi Q} e^{i s^i R_i^{\hphantom{i}j}\left(-\psi\right) S_j}.
\end{equation}
Once this is passed through the remaining terms, we find
\begin{equation}
e^{is^i S_i} U\left(x, \psi, \sigma^i, \chi, \theta\right) = e^{ix^\mu P_\mu} e^{i\psi Q} e^{i\left(\sigma^i + s^j R_j^{\hphantom{j}i}\left(-\psi\right)\right)S_j} e^{i\chi D} e^{i\theta J}.
\end{equation}
Therefore, under shifts parametrized by $s^i$, everything remains the same except the shift NG candidates, whose transformations are
\begin{equation}
\label{eq:HSFShifttrans}
\sigma'^{\:i} = \sigma^i + R^i_{\hphantom{i}j}\left(\pi + kx\right) s^j.
\end{equation}

Under dilations, the computation proceeds exactly as in \autoref{eq:MFUnderD}, yielding
\begin{equation}
e^{i\lambda D} U\left(x, \psi, \sigma^i, \chi, \theta\right) = e^{i e^{-\lambda} x^\mu P_\mu} e^{i\psi Q} e^{ie^{\Delta \lambda}\sigma^i S_i} e^{i\left(\chi + \lambda\right)D} e^{i\theta J}.
\end{equation}
As a result, the coordinates, shift NG candidates, and dilation NG candidate transform:
\begin{equation}
\label{eq:HSFDiltrans}
x'^\mu = e^{-\lambda} x^\mu, \qquad \sigma'^{\:i} = e^{\Delta \lambda} \sigma^i, \qquad \text{and} \qquad \chi' = \chi + \lambda.
\end{equation}

Finally, we have rotations. Since $J$ commutes with everything except the translation generator, we find that the coordinates transform as in \autoref{eq:MFRottrans}. However, this time shifts commute with rotations and we have a rotation NG candidate, so under rotations parametrized by $\varphi$,
\begin{equation}
\label{eq:HSFRottrans}
x'^i = R^i_{\hphantom{i}j}\left(-\varphi\right) x^j \qquad \text{and} \qquad \theta' = \theta + \varphi.
\end{equation}

The last transformation we must consider is unbroken translations, for which \autoref{eq:OGHSFparametrization} is the better parametrization to use,
\begin{equation}
U\left(x, \pi, \sigma^i, \chi, \theta\right) = e^{i x^\mu \bar{P}_\mu} e^{i\pi Q} e^{i \sigma^i S_i} e^{i\chi D} e^{i\theta J}.
\end{equation}
Under unbroken translations, only the first exponential is affected, meaning only the coordinates change
\begin{equation}
\label{eq:HSFTrans}
x'^\mu = x^\mu + a^\mu.
\end{equation}

As in the metafluid subsection, the transformation laws can also let us write down transformations on the generators themselves, which is helpful in computing the Maurer-Cartan form:
\begin{align}
\label{eq:HSFMCformhelp}
e^{-i\theta J} e^{-i \chi D} P_\mu e^{i\chi D} e^{i\theta J} &= e^{\chi} R_\mu^{\hphantom{\mu}\nu}\left(-\theta\right)P_\nu,\\
e^{-i\chi D} e^{-i\sigma^i S_i} Q e^{i\sigma^i S_i} e^{i\chi D} &= Q + e^{-\Delta \chi} \sigma^i \epsilon_i^{\hphantom{i}j} S_j,\\
e^{-i\chi D} S_i e^{i\chi D} &= e^{-\Delta \chi} S_i.
\end{align}

To summarize, the nontrivial transformation laws of the coordinates and fields under the full symmetry group are given by \autoref{eq:HSFU1trans}, \autoref{eq:HSFShifttrans}, \autoref{eq:HSFDiltrans}, \autoref{eq:HSFRottrans}, and \autoref{eq:HSFTrans}.

\bibliography{main}

\end{document}